\documentclass[sigconf]{acmart}

\usepackage[normalem]{ulem}

\usepackage[tableposition = top]{caption}
\usepackage{amsmath}
\usepackage{mathtools}

\usepackage{IEEEtrantools}
\usepackage{mathptmx} 
\usepackage{graphicx}
\usepackage{pbox}
\usepackage{circuitikz}
\usepackage{multirow}
\usepackage{float}
\usepackage{algorithm2e}
\usepackage{booktabs}

\usepackage{booktabs}
\usepackage{svg}
\usepackage{threeparttable}
\usepackage{tabularx}
\usepackage{pifont}
\usepackage{array}
\usepackage{lmodern}

\newcommand{\specialcell}[1]{\begin{tabular}[c]{@{}c@{}}#1\end{tabular}}
\newcolumntype{L}[1]{>{\raggedright\arraybackslash}p{#1}}
\newcolumntype{C}[1]{>{\centering\arraybackslash}p{#1}}

\input{Qcircuit}

\begin{document}
\setcopyright{rightsretained}
\copyrightyear{2017} 
\acmYear{2017} 
\setcopyright{acmlicensed}
\acmConference{MICRO-50}{October 14--18, 2017}{Cambridge, MA, USA}\acmPrice{15.00}\acmDOI{10.1145/3123939.3123949}
\acmISBN{978-1-4503-4952-9/17/10}

\title[Optimized Surface Code Communication in Quantum Computers]{Optimized Surface Code Communication in Superconducting Quantum Computers} 
\author{Ali Javadi-Abhari}
\affiliation{
  \institution{Princeton University}
}
\email{ajavadia@princeton.edu}

\author{Pranav Gokhale}
\affiliation{
  \institution{University of Chicago}
}
\email{pranavgokhale@uchicago.edu}

\author{Adam Holmes}
\affiliation{
  \institution{University of Chicago}
}
\email{adholmes@uchicago.edu}

\author{Diana Franklin}
\affiliation{
  \institution{University of Chicago}
}
\email{dmfranklin@cs.uchicago.edu}

\author{Kenneth R. Brown}
\affiliation{
  \institution{Georgia Institute of Technology}
}
\email{ken.brown@chemistry.gatech.edu}

\author{Margaret Martonosi}
\affiliation{
  \institution{Princeton University}
}
\email{mrm@princeton.edu}

\author{Frederic T. Chong}
\affiliation{
  \institution{University of Chicago}
}
\email{chong@cs.uchicago.edu}

\renewcommand{\shortauthors}{A. Javadi-Abhari et al.}


\begin{abstract} 
Quantum computing (QC) is at the cusp of a revolution.  Machines with 100 quantum bits (qubits) are anticipated to be operational by 2020~\cite{googlemachine,gambetta2015building}, and several-hundred-qubit machines are around the corner. Machines of this scale have the capacity to demonstrate {\em quantum supremacy}, the tipping point where QC is faster than the fastest classical alternative for a particular problem.  Because error correction techniques will be central to QC and will be the most expensive component of quantum computation, choosing the lowest-overhead error correction scheme is critical to overall QC success. This paper evaluates two established quantum error correction codes---planar and double-defect surface codes---using a set of compilation, scheduling and network simulation tools. In considering scalable methods for optimizing both codes, we do so in the context of a full microarchitectural and compiler analysis. Contrary to previous predictions, we find that the simpler planar codes are sometimes more favorable for implementation on superconducting quantum computers, especially under conditions of high communication congestion.
\end{abstract}



\maketitle

\section{Introduction}\label{sec:Introduction}
Major academic and industry efforts are underway to build scalable quantum computers, which can have significant applications in solving currently intractable problems in areas as diverse as AI~\cite{HHL}, medicine~\cite{lidar_chem} and security~\cite{Shor}. They will also profoundly impact our understanding of the nature of computation itself~\cite{aaronson2005}.

Owing to considerable theoretical successes over the past two decades~\cite{Threshold_Theorem,FT,kitaev2003fault,FowlerSurface} and successful small-scale physical demonstrations~\cite{kelly2015state,takita2016demonstration}, the main challenges now relate to scaling. This includes building more and longer-lived qubits, while simultaneously reducing the number of qubits and time required to run a given application. This paper focuses on the latter. We propose optimizations that reduce the number of qubits and clock cycles (the space-time resource usage), especially due to qubit communication requirements. We then discuss criteria that should be considered in choosing the lowest-overhead error correcting code.

Our focus on quantum error correction (QEC) stems from its importance in large-scale QC. Although it is expected that computers with 50-100 qubits will be built in a few years (and surpass classical computing capabilities due to exponentially increasing information state space), they will likely not sustain lengthy computations. This is due to the fragility of quantum states and the imprecisions of quantum control. Thus, in the long term, QEC is a necessity. However, QEC is the most resource-consuming component of QC, predicted to utilize up to 90\% of qubits in the system~\cite{isailovic2008running,Jones}. Any design should therefore aggressively optimize QEC, and also choose the best code among different possible schemes~\cite{Concatenated,FowlerSurface,dennis2002topological}. This paper hints at how this may be achieved.

In this paper, we study the most promising proposal for building large-scale quantum computers: {\em surface code} QEC on {\em superconducting} quantum technology~\cite{divincenzo2009fault}. These are being targeted in most industry efforts~\cite{googlemachine,gambetta2015building,rigetti2016}. Superconducting technology relies on qubits created from electrical circuits cooled to very low temperatures. They have the advantage of being both low error and fast. Surface codes are a family of codes that protect information using a simple 2-dimensional layout of redundant qubits with desirable scaling properties and high error resilience. 
Our study quantitatively compares two variations of the surface code in the context of a comprehensive microarchitectural and compiler analysis.  Considering issues like locality, concurrency, and congestion, our evaluation is informed by classical microarchitectural techniques, while bringing quantitative clarity to the most important QC design decisions.

The key challenge of this paper is to frame quantum computing problems, which
look very different from traditional architecture problems, in a manner that allows
us to apply computer architecture techniques. This application-system-technology co-design 
is crucial if the scarce quantum resources are to be used effectively. In particular:

\begin{enumerate}
  \item Surface code error correction has traditionally involved statically scheduling ``braids'' by hand in a 3D space-time volume for very small quantum circuits~\cite{Fowler12,mequanics}. Leveraging the property that program inputs are often fully-determined in QC applications, we map the problem from a set of 3D topological transformations to a 2D static routing problem, over-constraining it to allow scalability to very large circuits. We show that using proper heuristics, this approximation can still achieve near-optimal performance.
 
  \item Surface code ``braids'' have been greatly favored in the quantum computing community because they appear latency insensitive (an entire braid can be implemented in 1 cycle, regardless of length).  Yet we find that highly parallel quantum programs scale poorly because simultaneous braids can neither cross nor be prefetched, causing a form of contention scaling not previously analyzed. We show that in such cases, ``teleportation-based'' communication is more desirable as it is prefetchable. (Quantum resources can be pre-communicated, allowing any contention to be diffused over time prior to the point of use.)
 
  \item We incorporate (1) and (2) into an extensive software toolchain that performs end-to-end synthesis from high-level quantum programs to physical layout and circuits.  This automated framework allows us to map the design space parameterized by application characteristics, error correction scheme, and device reliability.  We show the importance of (2) by plotting optimal space-time design points as device reliability evolves with better technology. 
  
\end{enumerate}

The rest of this paper is organized as follows:
Section~\ref{sec:background} is an overview of the relevant background, and Section~\ref{sec:related} discusses related prior work. Sections~\ref{sec:microarch} and~\ref{sec:toolflow} discuss the micro-architectural details of various QEC implementations, and the software stack designed for evaluations of the design space. Section \ref{sec:braid} discusses optimizations on braid-based communications, and Section~\ref{sec:results} shows our results. Section~\ref{sec:discuss} offers broader discussions including options for other communication optimizations, and Section~\ref{sec:conclusion} concludes the paper.

\section{Background}\label{sec:background}
This section presents a brief overview of the relevant background on quantum computation, theory of error correction, and physical technologies.

\subsection{Principles of Quantum Computation}
Quantum bits ({\em qubits}) are the principal units of information in quantum computers. Unlike classical bits, their states are not confined to two deterministic binary values. Instead, a generic quantum state exists in a probabilistic yet simultaneous superposition of the $\ket{0}$ and $\ket{1}$ states. Manipulation of probabilities is equivalent to changing the state, and can be achieved through the application of various quantum {\em operations}. When observed (measured), a qubit collapses into a classical binary state (either 0 or 1), in accordance with the prior probabilities of the quantum state. Expressed in algebraic terms, quantum states are unit vectors in the Hilbert space. Quantum operations rotate these vectors, while quantum measurements project them onto a lower dimension. 

While this means that there are theoretically infinite possible quantum states, and infinite valid operations to transition between those states (any small degree of rotation is permissible), it can be proven that a small set of operations is sufficient to approximate all possible operations~\cite{kitaev1997quantum,Barenco}. These {\em universal operations} are akin to a classical instruction set. 

The power of quantum computation comes from its exponentially increasing state space: $n$ qubits are represented by a $2^n$-dimensional state vector. Furthermore, quantum states can interfere, owing to their dual nature as particles and waves, further allowing manipulations that surpass the power of classical computers. The key idea behind designing quantum applications is to initialize qubit states to encode a given problem, orchestrate interferences to eliminate undesired answers from the state space, and finally measure the qubits, obtaining (with high probability) the answer to the query.

\subsection{Theory of Quantum Error Correction}
When a quantum application is designed, it assumes noise-free computation. In practice, however, the computation can deviate from the intended path due to noise, interaction of qubits with the environment, and imprecise operations. Fortunately, if the noise level is within specific bounds, this deviation can be corrected through constant monitoring and error detection~\cite{Threshold_Theorem}.

As with any type of error resilience, redundant information is used to protect fragile states. Although we cannot simply copy the state of one qubit into multiple qubits and later take a majority vote (due to quantum no-cloning constraints~\cite{NoCloning}), we can still use redundant qubits to create a code space where valid code words are safely separated, immune to noise perturbations. The idea is similar to classical error correction: more redundancy will result in a larger {\em code distance (d)}, increasing error tolerance. Small corruptions of data would place the data's state into an orthogonal error subspace, which can be immediately detected and corrected.
The higher-level encoded quantum state is called the {\em logical qubit}, while its comprising fundamental qubits (including redundant bits) are called {\em physical qubits}.

One complicating factor in checking logical qubits for errors is that we cannot directly observe their comprising physical qubits---observing a qubit collapses its state superposition, yielding a simple classical bit. This necessitates the use of extra ``helper'' qubits, known as {\em ancillas}.  Ancillas are interacted with data for {\em syndrome measurement}---learning just enough about the encoded block to detect and fix errors without measuring the full state of the encoded block. Even though quantum states are analog, and errors can occur in arbitrary amounts, syndrome measurement has the effect of projecting the qubit state onto the basis vectors of the error subspace, thereby quantizing errors.  Such quantized errors can then be corrected using simple corrective operations: a bit-flip operation, a phase-flip operation, or both.

After encoding, any computation must be performed directly on encoded data, since decoding and encoding again would be too error prone. Performing encoded versions of some operations are harder than others. In particular, most proposals for performing the T operation require the use of even more extra qubits. The extra qubits must first be prepared into one specially encoded quantum state (called ``magic state'' due to its distinctive properties~\cite{MikenIke,fowler2009high,magic_states}), and then interacted with the original data to perform the T operation.

Unfortunately, the work required to achieve proper error resilience is quite costly, since there is a wide gap between reliability requirements of non-trivial quantum applications and reliability rates supported by quantum technologies (called {\em logical error rates} ($p_L$) and {\em physical error rates} ($p_P$) respectively). It is typical to have $10 - 15$ orders of magnitude difference between $p_L$ and $p_P$. The application imposes a far more stringent error-per-operation expectation than the technology can support.
$p_L$ and $p_P$ are dictated by the {\em size of computation} (i.e. total number of pre-QEC logical operations that must be executed) and the noise level in the physical technology, respectively. 
Longer computation has a higher chance of corruption, thus requiring more reliable operations. For example, an application that executes a total of $10^{12}$ logical operations must ensure that per-operation errors do not exceed $0.5\times10^{-12}$, if it wants to accurately perform its computation at least half the times it is executed. 50\% is a typical correctness target, which we also assume. 
On the other hand, current superconducting technology is able to perform operations with reliabilities of $99.99 - 99.999\%$---equivalent to physical error rates of $10^{-2} - 10^{-3}$. 
A main goal of this paper is to optimize QEC in a way that we do not lose more resources to it than necessary. In addition, we also suggest the least resource-intensive QEC code for a range of logical-to-physical error gaps, as this parameter looks very different for various applications and technologies.

\subsection{Surface Codes}\label{subsec:sc}
\begin{figure}[tb]
\centering
\subfloat[Planar\label{fig:surface_planar}]{
  \includegraphics[width=0.12\textwidth]{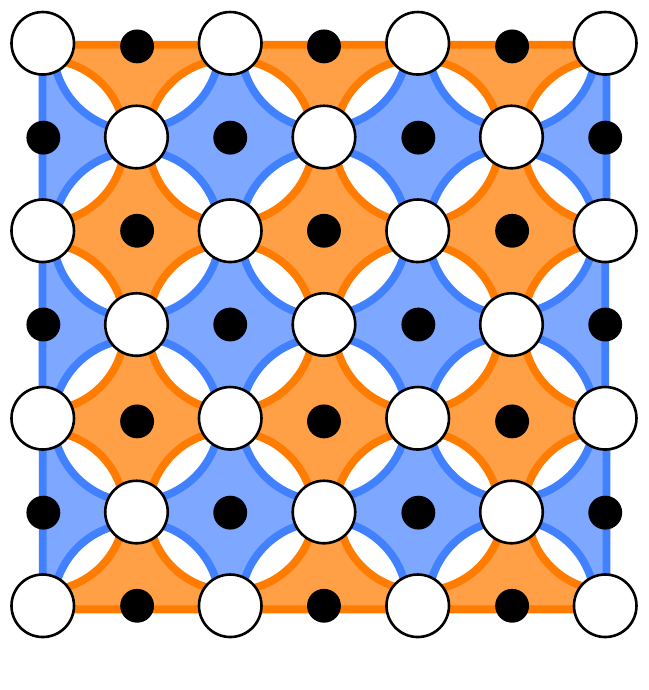}
}
\subfloat[Double-Defect\label{fig:surface_dd}]{
  \includegraphics[width=0.27\textwidth]{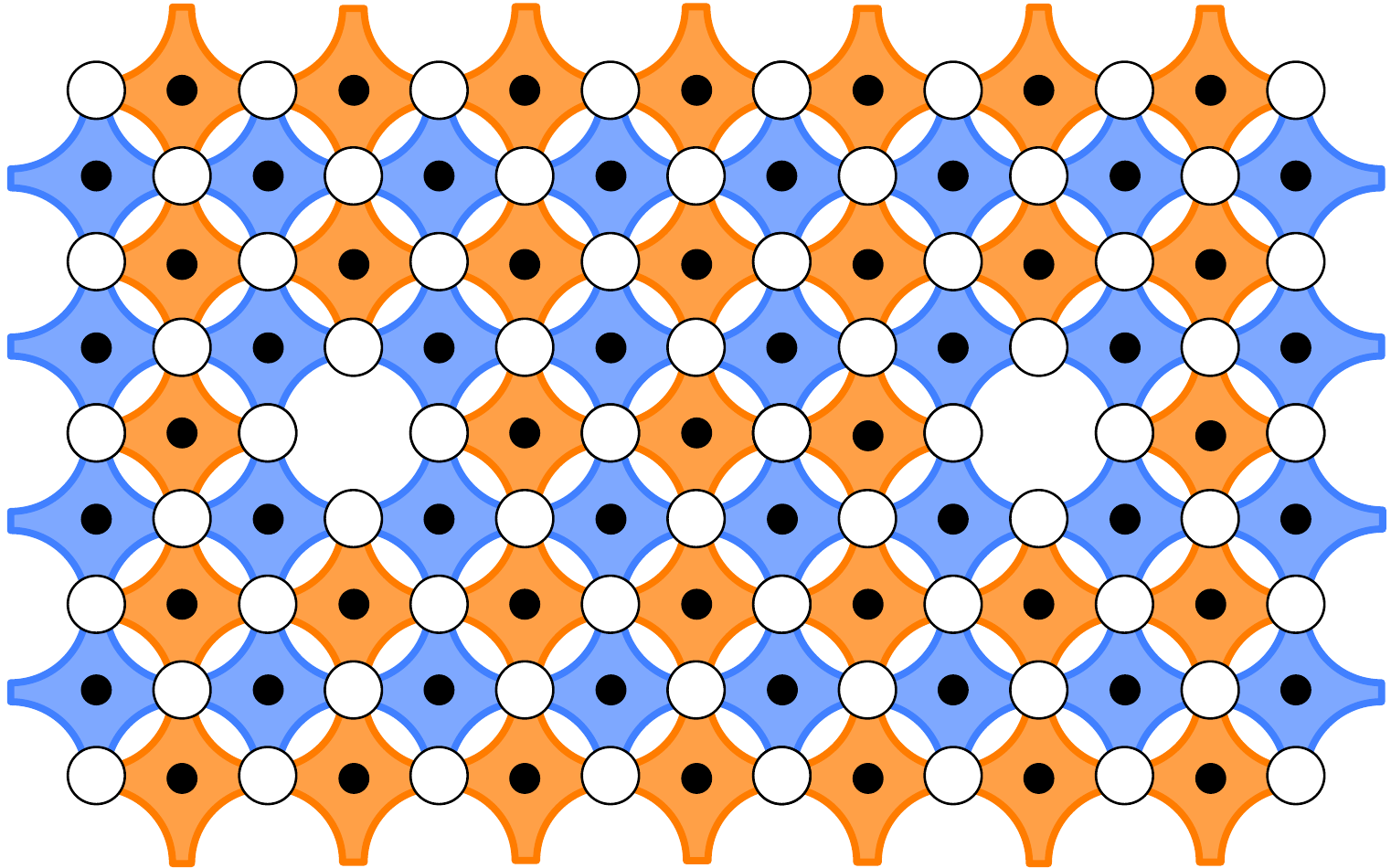}
}
\caption{One encoded logical qubit, using the (a) planar and (b) double-defect surface code variations. Ancilla qubits (small black) are used to continuously monitor data qubits (large white). The planar encoding uses fewer physical qubits for the same encoding strength.}
\vspace{-10pt}
\label{fig:surface}
\end{figure}

Surface codes are a family of promising, low-overhead QEC codes~\cite{FowlerSurface}, where logical qubit information is encoded in the topology of a two-dimensional lattice of physical qubits~\cite{RaussendorfSurface}. 
A surface code lattice alternates data and ancilla qubits (white and black circles in Figure~\ref{fig:surface}). Data qubits collectively encode the logical state, while ancillas continuously collect syndrome measurements from their neighboring data. It can be shown that the lattice as a whole acts as an encoded qubit. A larger lattice is a stronger encoding (larger code distance $d$).

Intuitively, surface codes are so fault tolerant because they employ a ``soft decoding'' scheme in which error syndromes are measured and recorded over an extended time period, and then a minimum-weight perfect-matching algorithm~\cite{edmonds1965maximum} is used to identify errors after the fact (and then the errors are post-corrected due to commutativity of the correction operations).

\subsubsection{Planar vs. Double-Defect Encodings}
In this paper we investigate the two main flavors of surface codes: {\em planar}~\cite{bravyi1998quantum,dennis2002topological} and {\em double-defect}~\cite{FowlerSurface} encodings. In the planar encoding (Figure~\ref{fig:surface_planar}), a single lattice (or ``plane'') represents a single logical qubit. To introduce extra logical qubits to the system, it suffices to build separate lattices. In this scenario, the system can be seen as a collection of planar tiles. In contrast, the double-defect implementation (Figure~\ref{fig:surface_dd}) introduces holes (or ``defects'') into the lattice, which are locations where the syndrome measurements are turned off. This creates a standalone qubit between the two defects (the lattice boundary is no longer needed), and thus a monolithic lattice is now able to accommodate multiple connected double-defect tiles of this type.

\begin{figure}[t]
\centering
\subfloat[Planar\label{fig:syndromes1}]{
  \includegraphics[width=0.46\columnwidth]{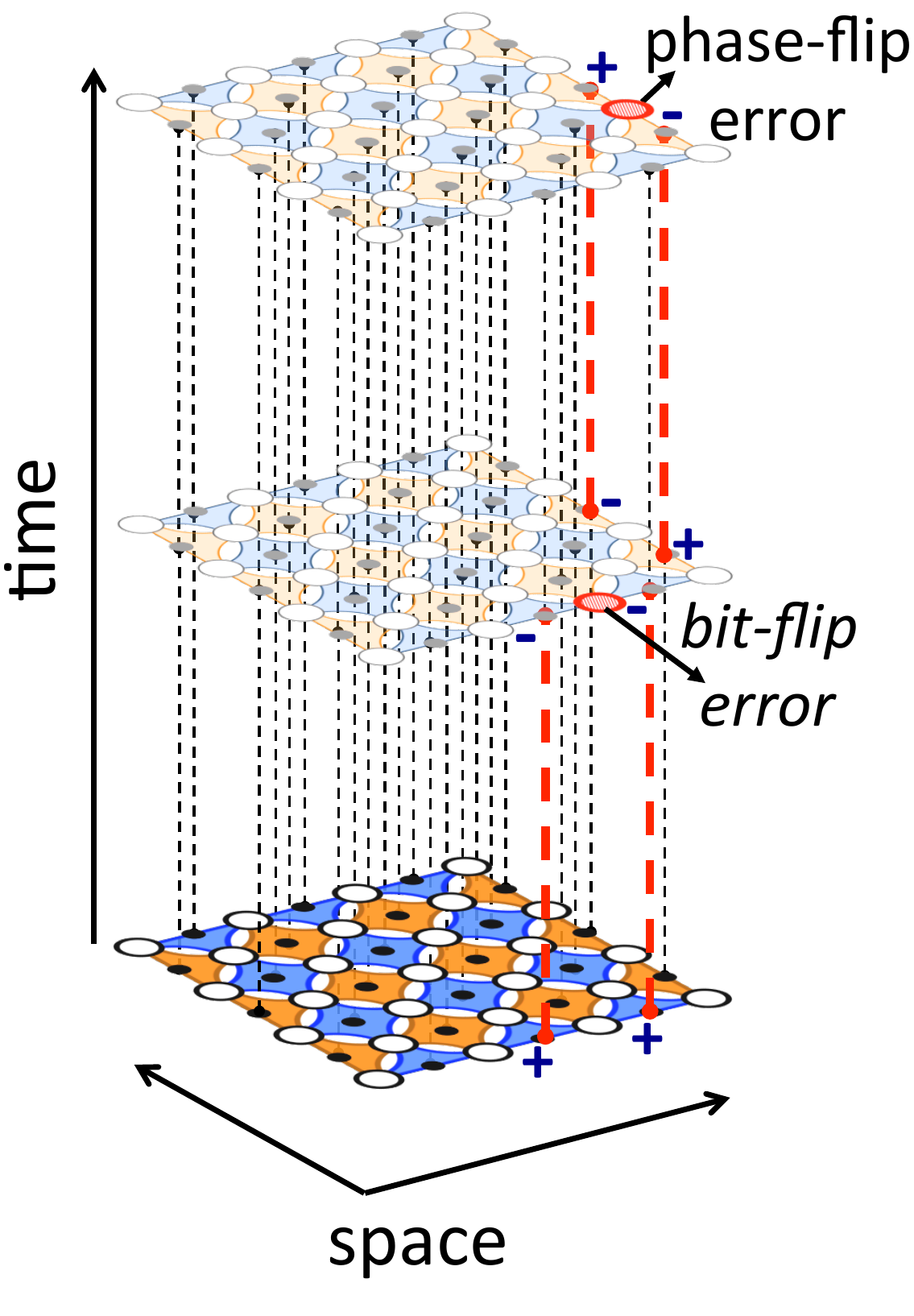}
}
\subfloat[Double-Defect\label{fig:syndromes2}]{
  \includegraphics[width=0.54\columnwidth]{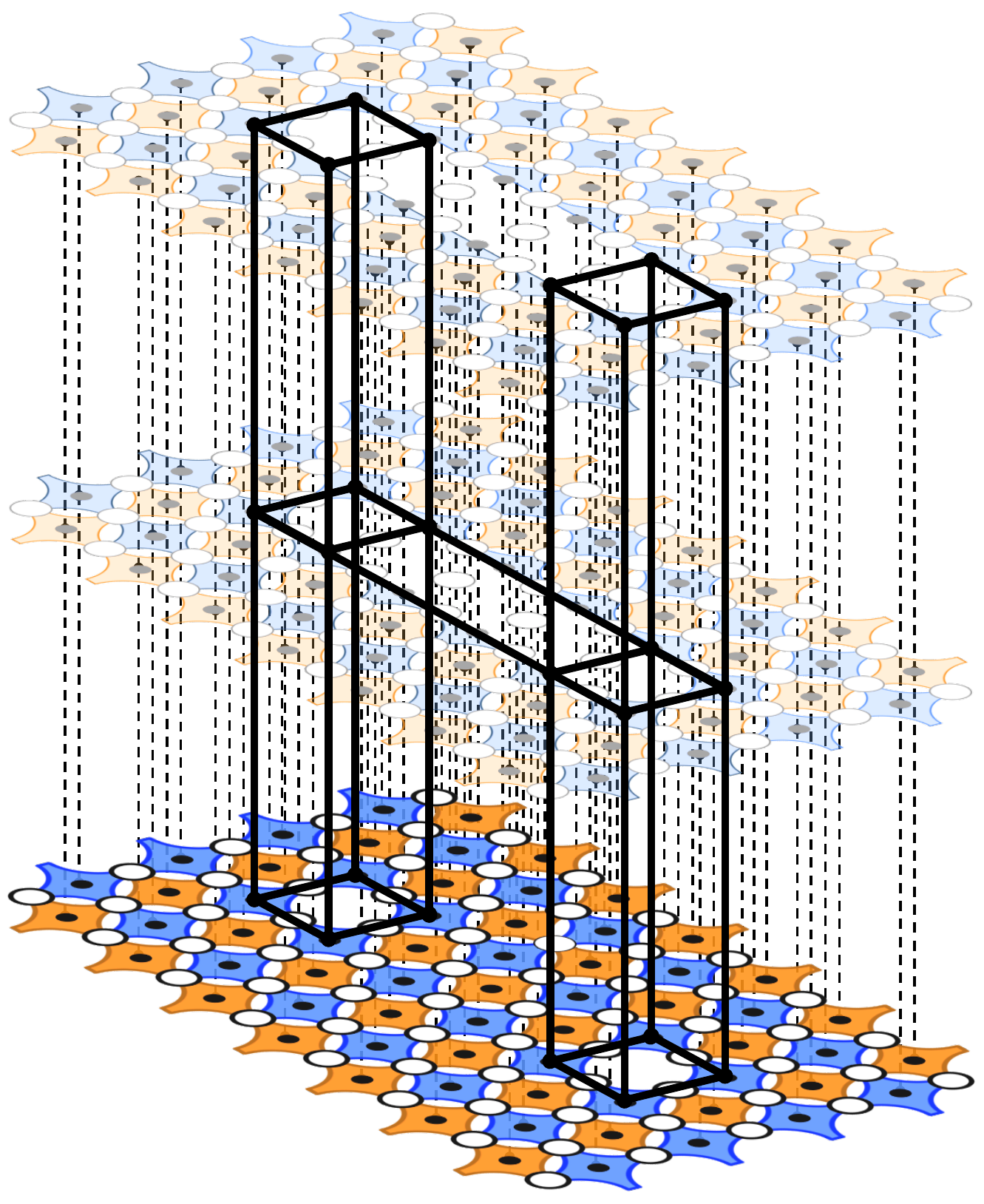}
}
\caption{(a) Surface code error detection. A matching algorithm over anomalous syndromes can determine the location of faulty qubits. (b) Defects that are stretched through space-time create ``braids'' --- 3D pipes used to operate on the encoded logical qubit.}
 \vspace{-14pt}
  \label{fig:syndromes}
\end{figure}

As discussed previously, detecting bit-flips and phase-flips is sufficient to correct arbitrary qubit errors. Therefore, ancillas are alternately designated as those that measure bit-flip syndromes and those that measure phase-flip syndromes from their surrounding data (shown as orange and blue interactions in Figure~\ref{fig:surface}). By recording a history of syndrome measurements, we are able to detect (in classical software) anomalous syndromes and trace their cause to specific qubit corruptions. This collection of syndrome histories can be thought of as a 3D space-time volume; the bottom area being the lattice, and the vertical height the progress in time. This is shown in Figure~\ref{fig:syndromes}. 

\subsection{Superconducting Technology}
A number of technologies have been proposed for implementing qubits. Ongoing experimental physics research seeks to create a technology that can accommodate large numbers of stable qubits, operations with high fidelity, and the ability to easily address and control the system's many qubits. In this work we focus on superconducting technology as one promising step towards those goals. Owing to fast clock speeds and low error rates~\cite{chow2012universal,hifi_supercond_1,hifi_supercond_2}, superconducting qubits have become the leading candidate for large-scale QC. This technology has a good prospect of scaling up to a large number of qubits~\cite{SuperCondOutlook}, and operation clock rates are in the high range of quantum technology, currently on the order of 10-100 MHz~\cite{paik2011observation,wallraff2004strong}. 
The qubits in this technology are difficult to move, and thus only interact over short distances with nearby qubits. The implications of this on qubit communication is discussed in Section 4. Superconducting systems meeting the necessary criteria to run surface codes have already been demonstrated~\cite{hifi_supercond_1}.

\section{Related Work}\label{sec:related}
Several prior works have estimated quantum resources using simple tallying of logical qubits and operations, and multiplying them by the number of resources necessary to correct one qubit and one operation~\cite{Jones,Suchara,FowlerSurface}. This ``spreadsheet'' model simplifies several crucial constraints---for example, that distant qubits cannot arbitrarily interact. In this work we perform a more rigorous evaluation of the details of mapping to the architecture, taking into account data dependencies, qubit layouts, and contentions in communication. This exposes new opportunities; for example we find that communication-aware scheduling saves up to ${\sim}7X$ in total execution time by reducing braid congestion.

The initial assignment of the program's logical qubits to the chip's physical qubits is an important optimization which can positively impact performance, if done correctly. Prior work has considered this problem, although several shortcomings exist. For example, some methods are manual or non-scalable, some only consider 1-dimensional architectures, and some ignore error correction or only consider concatenated error correction~\cite{whitney2007automated, saeedi2011synthesis, chakrabarti2011linear, choi2011effect, pham20132d, dousti2012minimizing, shafaei2013optimization, shafaei2014qubit, lin2014ftqls, lin2015paqcs, pedram2016layout}. For the first time, we optimize two-dimensional surface code mappings by applying graph partitioning techniques to quantum programs to both achieve scalability and to reduce communication distance and congestion.

Furthermore, we examine the sensitivity of results to the characteristics of the high-level application and reliability of the low-level technology, parameters which evolve over time. While prior work has often relied on optimizing a particular QEC/technology choice~\cite{cqla,FowlerSurface,Jones}, our work provides insights into the conditions where such choices are warranted. In a rapidly-moving field, this end-to-end toolflow is particularly important since it allows one to change approaches fluidly to try new methods as technologies or assumptions evolve.

Double-defect codes have been the favored QEC option in the literature~\cite{FowlerSurface,Jones,fowler2012bridge,paler2016synthesis}. This is due to simplicity of its uniform control and the fact that a monolithic lattice can in theory be made as large as needed. In addition, this encoding uses braids for communication, which are very fast forms of interacting distant qubits. However, we find that under certain conditions planar codes are actually better, owing to two properties: first, planar tiles are smaller (i.e. fewer qubits needed for the same code distance). Second, planar tiles communicate through {\em teleportations}, which are prefetchable (a major advantage in high-contention scenarios).

Recent work has begun addressing the issue of braiding automation by performing automated constructions of 3D topological pipes~\cite{paetznick2013quantum,paler2015compiler,paler2016synthesis}. However scalability and extending to all operations remains an area of future work in these methods. Our proposed alternative is to use more scalable 2D networking methods. Several recent papers have proposed optimized qubit layouts on 2D grids~\cite{shafaei2013optimization,dousti2014squash,lin2014ftqls}. However, this problem has not been studied in the context of surface codes or in larger scales. Our work fills this gap.

Finally, this work demonstrates the importance of paying attention to the high-level application when choosing the best type of error correction, a factor that has received little attention in prior work. Patil et al.'s study~\cite{patil2014characterizing} similarly reveals that high-level application characteristics must be taken into account when choosing the right implementation of a quantum library subroutine, in the event that many implementations of it are available.

\section{Microarchitectural Design}\label{sec:microarch}

In this section, we describe the microarchitectural components necessary to support computation on qubits encoded in the planar and double-defect schemes. In this section, we focus on logical qubits and in later sections will implement these logical qubits with multiple physical qubits in our evaluation.
We begin by discussing several important factors that must guide microarchitectural design, and then discuss our evaluated microarchitectures.

\subsection{Handling Data Movement}\label{subsec:movement}
A requirement of universal quantum computing is long-range interaction of logical qubits. First, because an algorithm needs to operate on physically-remote qubits. Second, this is required since magic states are prepared in dedicated regions of the hardware, yet need to be communicated to the point of use when a T operation is needed. Many technologies, including superconductors, do not allow easy transport of qubits. However, two methods exist to communicate the data in surface-encoded logical qubits: {\em teleportation} for planar qubits~\cite{teleportation}, and {\em braiding} for double-defect qubits~\cite{FowlerSurface}.

Teleportation is a technique for long-distance communication of exact quantum states. It works by qubit {\it entanglement}, which causes operations on one qubit to affect the state of another entangled qubit, even if the two are no longer physically close. When the source and destination qubits of the communication are both entangled to an intermediary qubit, they are ``linked,'' even when far apart. 
However, this does not entirely solve the problem of data movement. In order to establish a virtual link, two qubits can become entangled while physically together, but they must then be physically transported to the desired source and destination locations of the communication. This is known as {\em EPR distribution}---supplying pairs of qubits placed in a special EPR state~\cite{teleportation,Bell,MikenIke}. Fortunately, since EPR pairs are independent of data, they may be entangled and distributed (i.e. prefetched) in advance and in fairly latency-tolerant manner.

{\em Swapping} is another suitable method for physical movement of quantum data from one location to another. Since superconducting qubits may only interact with close-range qubits, a chain of swaps can be used to achieve the same effect as long-distance mobility. Swapping is an expensive operation. Not only does the travel time increase substantially with distance, but the channels that are used for movement also consume extra ``dummy'' qubits, who exist only for the purpose of facilitating swaps. However, once EPRs are distributed, teleportation only has a small constant latency, independent of distance. The parameters of swaps and teleportations have been derived in~\cite{oskin2003building}.

While planar codes can use teleportation to communicate, double-defect codes use an alternate communication form known as {\em braiding}. Braiding consists of extending defects through space (turning off all syndrome measurements along the way). The source and destination of the braid will become entangled, and the braid can then shrink back to its original starting defect. The advantage of braids is that their extension and shrinkage only takes one cycle, regardless of the distance. Their disadvantage, however, is that they cannot cross, and they cannot be prefetched---each communication must occur at the point it is needed since there are no separate communication ancillas involved. To summarize, Table~\ref{tab:comms} shows the main differences in how the two communication methods behave.

\begin{table}[t]
  \footnotesize
  \centering
  \caption{Summary of tradeoffs in communication efficency among the two main flavors of the surface code.}
  \renewcommand{\arraystretch}{1.3}
  \setlength{\tabcolsep}{1pt}  
  \begin{tabular}{>{\raggedright\bfseries}m{0.5in} >{\centering}m{0.85in} >{\centering}m{0.6in} >{\centering}m{0.6in} >{\centering\arraybackslash}m{0.6in}}
    \toprule
    & \bf{\specialcell{Communication\\Method}} & \bf{\specialcell{Space\\(Qubits)}} & \bf{\specialcell{Time\\(Latency)}} & \bf{\textbf{Pre-fetchable?}} \\
    \midrule    
    \bf{\specialcell{Planar}}                  & Teleportation 
                                               & Low
                                               & High
                                               & Yes \\
    \midrule
    \bf{\specialcell{Double-Defect}}         & Braiding 
                                               & High
                                               & Low 
                                               & No \\                            
    \bottomrule
  \end{tabular}
  \label{tab:comms}
  \vspace{-8pt}
\end{table}

\subsection{Differences to Classical Communication}
While the problem of communicating qubits may seem similar to a classical networking problem, in fact several key distinctions call for new approaches. First, quantum data is physical---it cannot exist in two places at once and cannot be copied. This puts more pressure on QEC since any corrupted or lost data cannot simply be retransmitted. Second, communication not only has a time overhead, it also incurs a space overhead in terms of ancillas. Third, quantum applications are almost always specialized to concrete problem inputs (e.g. factoring a 2048-bit number). This means that the execution trace and therefore the point of each communication is known in advance. This global knowledge yields opportunities for aggressive optimization.

More specifically, the two surface code communication options discussed present new challenges. 
Teleportation is a non-traditional 2-step communication process, where step 1 may cause collision but is prefetchable. The goal of the optimizer must be to orchestrate this highly flexible step. Namely, do not distribute EPRs too early since they may cause traffic, and do not distribute too late since they may stall computation. A smooth, low-contention ``just-in-time'' distribution can be achieved with full knowledge of the program path. 
In braiding, there is no actual transmission of information (only expansion of lattice defects). So a braid can stretch all the way in 1 error correction cycle. However, these defects cannot physically co-exist close by, prohibiting crossing, buffers or virtual channels. The static availability of communication points again facilitates optimization heuristics to reduce traffic.

\subsection{Ancilla Generation}~\label{subsec:ancilla}
In addition to the physical ancillas required for syndrome measurement on the surface code lattice, we also need ancillas at the logical level. In particular, a steady supply of magic states and EPR states are needed for performing T operations and teleportations, respectively.

We use so-called ``ancilla factories''~\cite{Jones,isailovic2008running,van2010distributed,steane1997space} to dedicate specialized regions of the architecture to continuously prepare and supply ancillas. This creates more communication pressure: for every T operation and every teleportation, the relative ancillas are produced in factories and consumed at the location of data. Furthermore, ancilla factories can be large---every magic state factory consumes 12 encoded qubits (and much more on a faultier device)~\cite{Jones}. Previous studies have found that a factory-to-data footprint of almost 10:1 is needed if ancilla generation is to be taken off of the critical path\cite{isailovic2008running}. Conversely, saving more space will cost more in time. In our empirical model, we have found that a good space-time balance is achieved with a 1:4 ancilla-to-data ratio.

\begin{figure*}[ht]
\centering
\hspace*{\fill}
\subfloat[Multi-SIMD Architecture\label{fig:microarch1}]{%
  \includegraphics[width=0.43\textwidth]{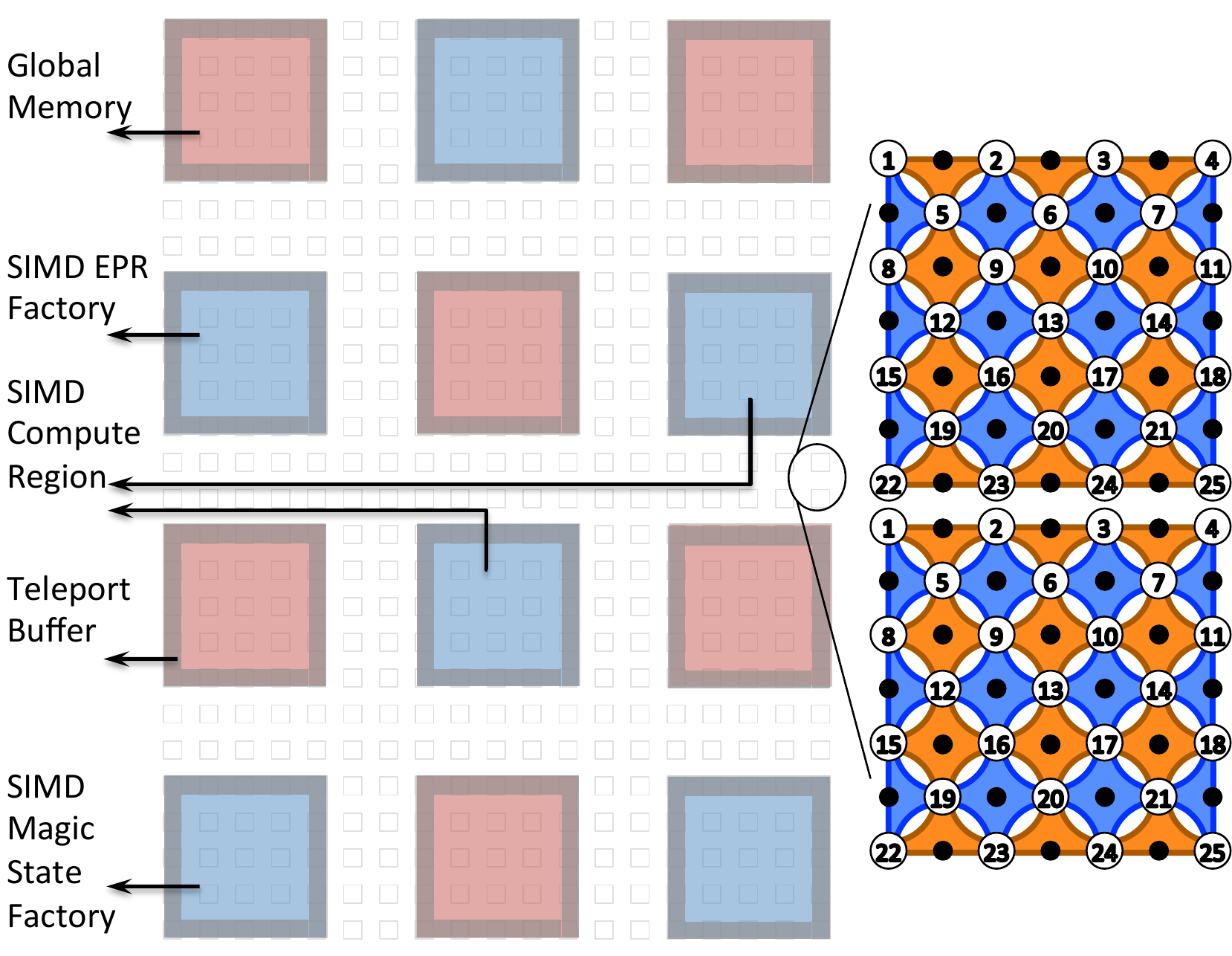}
}
\hfill
\subfloat[Tiled Architecture\label{fig:microarch2}]{%
  \includegraphics[width=0.39\textwidth]{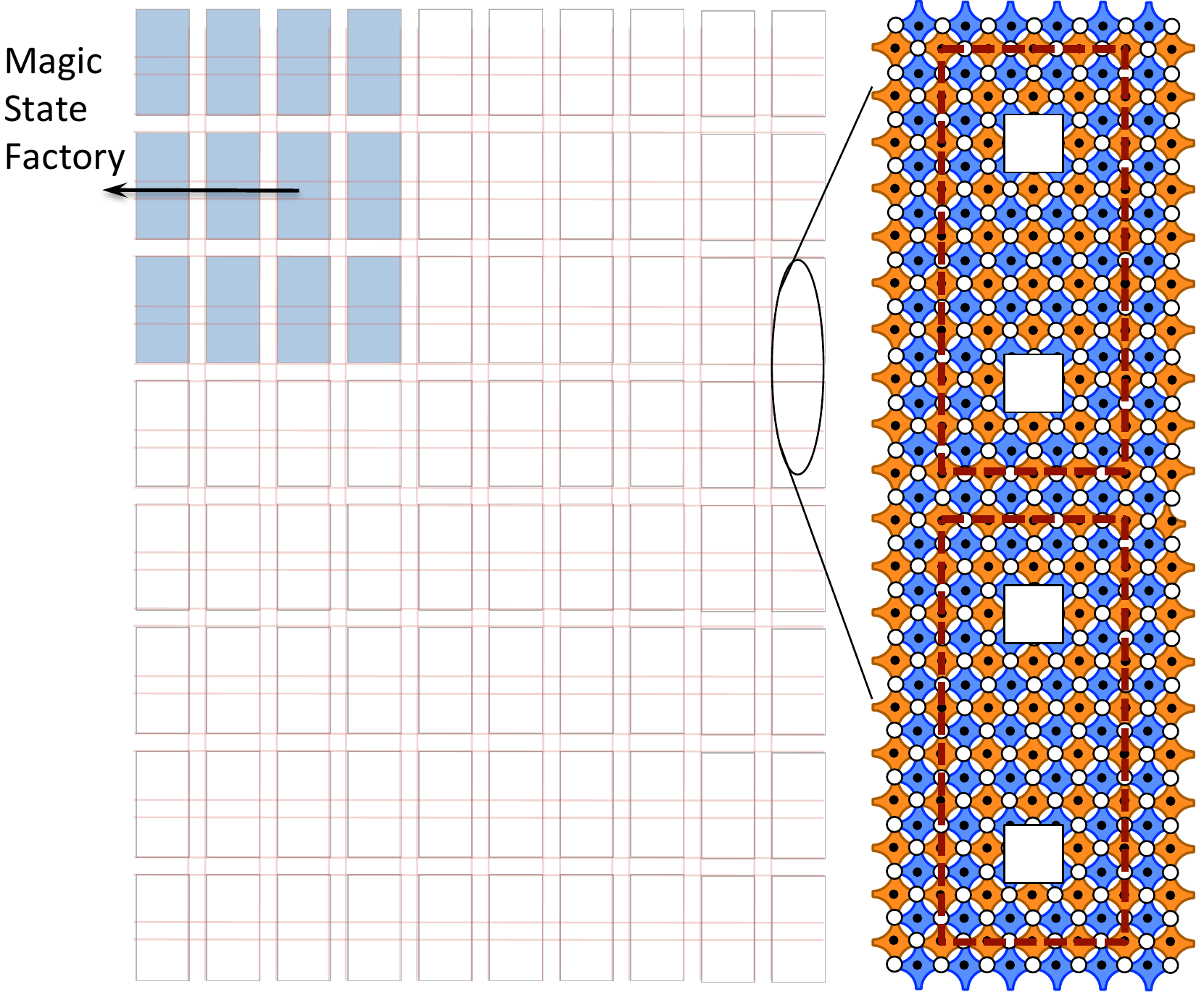}
}
\hspace*{\fill}
  \caption{Microarchitectural designs with zoomed-in physical layers. (a) Multi-SIMD architecture for planar QEC. SIMD regions (blue) are used to apply similar operations to many planar qubits in parallel. Distributed memory regions (red) store idle data. Dedicated SIMD regions are used as ancilla factories, for steady generation of magic states and EPRs. Each region is surrounded by a teleport buffer (grey), that entangle EPRs to data to be teleported. EPR distribution occurs through planar swaps---numbered qubits in the zoomed-in figure show bitwise couplings needed to perform swaps. (b) Tiled architecture for double-defect QEC. Braid channels (red) connect tiles (black), and also pass between the double defects in every tile. Dedicated factories supply magic states to surrounding tiles.}
\label{fig:microarch}
\end{figure*}

\subsection{Multi-SIMD Architecture for Planar QEC}
An advantage of planar codes, beside their smaller size, is that applying a logical operation on the plane amounts to applying identical bitwise operations on individual physical qubits inside the plane, and qubit communication can occur through bitwise interactions between two planes (Figure~\ref{fig:microarch1}). This suggests that a SIMD-style architecture is suitable. We use the {\em Multi-SIMD} architecture proposed by~\cite{heckey}, which combines qubit-level parallelism with operation-level parallelism: many qubits undergoing the same operation are clustered in one SIMD region, and multiple (reconfigurable) SIMD regions can accommodate heterogenous types of operations at any cycle. There is well-established technological support for this: microwave broadcasts can be used to operate simultaneously on many superconducting qubits~\cite{supercond_charge,supercond_flux}.

Figure~\ref{fig:microarch1} depicts the different regions at use: a checkerboard layout of SIMD and memory regions allows for easy access for qubit computation and storage. Movement is a natural consequence of this architecture. We use teleportation as it decouples the hard part of communication (the network congestion) from the data communication itself. The former is done on EPRs and the latter on data qubits. In this sense, only EPRs use the communication mesh, whereas data and magic states get teleported to where they are needed.

The bitwise coupling of planar qubits---needed for 2-qubit operations and swap channels---has traditionally been difficult in superconductors, because it requires 3D connections between non-adjacent qubits. Recent work in developing air-bridge crossovers~\cite{chen2014fabrication}, vias~\cite{hifi_supercond_1} and flip-chip crossovers~\cite{abraham2015removal} has, however, mitigated this problem. These connections can connect medium-distance qubits and are protected from cross-talk by shielding methods~\cite{brecht2015multilayer}. Also importantly, the development of low-loss, 3D connections seem to be necessary even in double-defect codes, because of the need for off-chip control and measurement. Other possible 3D architectures to couple medium-range planar qubits have also been described~\cite{helmer2009cavity}.

\subsection{Tiled Architecture for Double-Defect QEC}
The double-defect code defines all qubits on one large, monolithic lattice. The tiled architecture in Figure~\ref{fig:microarch2} assigns one tile per qubit, and opens channels between them to allow for communication braids. Similar to the Multi-SIMD architecture, we reserve some tiles for continuous generation of magic states, to be braided to various points of use. No EPR factory is needed, since no teleportation occurs.

\section{Toolflow and Applications}\label{sec:toolflow}
This section presents our comparative evaluation methodology, including the studied quantum applications and the end-to-end toolflow developed for this purpose.

\subsection{Overall Toolflow}
Designing a quantum computer is a complex task. Automated tools play an invaluable part in highlighting the tradeoffs of the design, and ensuring that every step is optimized such that quantum resources are used as efficiently as possible. Suppose, for example, that we wish to run a particular application on a given hardware. First, the application must be analyzed for its parallelism potential, and this must be matched with the microarchitecture and hardware's support for parallelism. Second, points of qubit communication during the program run must be identified, and the overhead due to network congestion accurately analyzed. Third, challenging space-time tradeoffs must be evaluated which go beyond mere qualitative reasoning. For example, in designing quantum applications it is typical to trade qubits for time~\cite{hastings2014improving}. However, these two parameters also affect each other. If we decrease qubits, time will be increased which causes more error accumulation, which may force us to compensate with stronger error correction which will in turn increase the qubits. In summary, a design toolflow must model all of these pieces, yielding a resource and performance estimate informed by application characteristics (e.g. parallelism), technology characteristics (e.g. clock speeds, reliability rates, mobility of qubits), and communication characteristics (e.g. contention, teleportation vs. braiding models, straight moves vs. turns). 

\begin{figure*}[tbp]
  \centering  
  \includegraphics[width=0.9\textwidth]{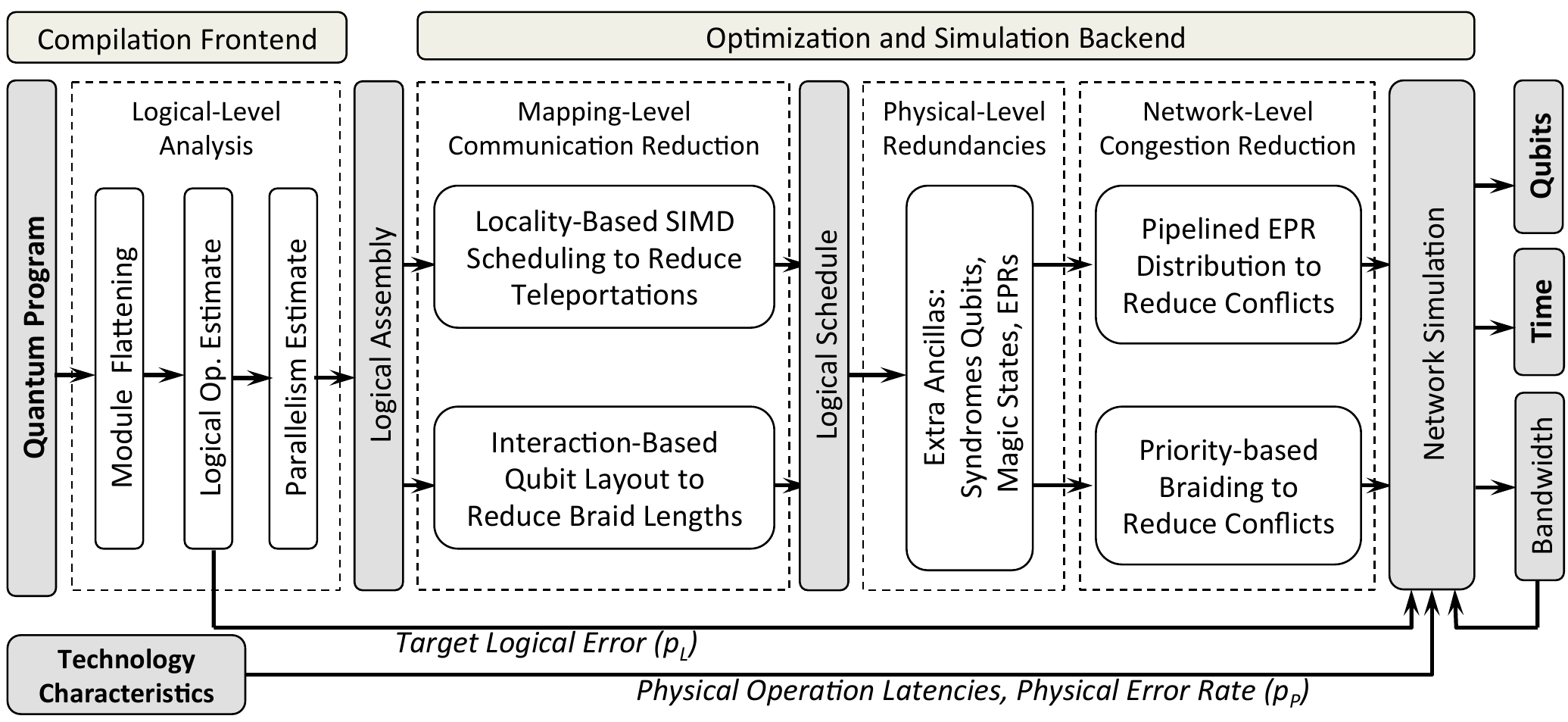}
  \caption{Overview of our toolflow. The frontend performs logical compilation, and is based on~\cite{ScaffCC}.
The backend performs mapping to and optimizations on the relevant microarchitecture.
The top optimization path pertains to teleportation-based communication in planar codes; the bottom path concerns braid-based communication in double-defect codes (Section~\ref{sec:braid}).}
  \label{fig:toolchain}
\end{figure*}

Our toolflow's integration allows for analysis of program graphs through multiple stages and application of various optimization heuristics. It also facilitates iterative designs through feedback capabilities. For example, the overhead of downstream error correction can often depend on the degree of code inlining in earlier compilation stages, which can be optimized iteratively (as discussed in Section~\ref{sec:results}).  Finally, the toolflow also enables sensitivity studies such as in Section~\ref{sec:sensitivity}). Figure~\ref{fig:toolchain} depicts our overall toolflow.

\subsection{Applications}
As input to our toolflow we use a set of applications of varying size, characteristics, and functions. Table~\ref{tab:apps} summarizes these. Of particular interest is the parallelism potential of applications, which affects the performance of optimization protocols, and therefore the tradeoffs between the different QEC and communication methods we consider. We evaluate two highly-parallel applications (IM and SHA-1) and two mostly-serial applications (SQ and GSE).

\begin{table}[t]
  \footnotesize
  \centering
  \caption{Summary of studied quantum applications. Parallelism factor is average number of logical operations that can be concurrently executed, were hardware resources not a constraint (ideal parallelizability).}
  \renewcommand{\arraystretch}{1.2}
  \setlength{\tabcolsep}{1pt}  
  \begin{tabular}{>{\raggedright\bfseries}m{1.05in} >{\centering}m{1.55in} >{\centering\arraybackslash}m{0.6in}}
    \toprule
                            & \textbf{Application Purpose} & \textbf{Parallelism Factor} \\
    \midrule
    Ground State Estimation (GSE)  & Compute ground state energy for molecule of size m~\cite{ref:gse} & $1.2$ \\
    \midrule    
    Square Root (SQ)  & Find square root of an n-bit number~\cite{Grover} & $1.5$ \\
    \midrule    
    SHA-1 Decryption (SHA-1)  & SHA-1 decryption of $n$-bit message~\cite{ref:sha1} & $29$ \\    
    \midrule    
    Ising Model (IM)  & Finding ground state for ising model on n-qubit spin chain~\cite{ref:im} & $66$ \\                                              
    \bottomrule
  \end{tabular}
\vspace{-10pt}  
  \label{tab:apps}
\end{table}

\subsection{Compilation Frontend}
We use the ScaffCC compiler~\cite{ScaffCC} as the frontend to our toolflow, lowering high-level descriptions of quantum algorithms to a standard logical-level ISA known as QASM (quantum assembly)~\cite{qasm,qasm2}.

The frontend performs logical-level resource and parallelism estimations. They guide the backend network optimization policy, as well as choice of code distance to meet the logical reliability requirements. That is, the resource estimation helps discover the number of logical operations that must be executed (size of computation), which is inversely proportional to the target logical error ($p_L$). This, in conjunction with the physical error rate ($p_P$) furnished by the technology characteristics, helps determine the strength of surface code error correction that is needed ($d$). 

\subsection{Optimization and Simulation Backend}
The backend is responsible for adding physical-level redundancies to ensure error tolerance, as well as performing optimizations and simulations to calculate the final space-time overhead. Optimizations occur at two levels: {\em mapping-level} reduction of total communication needs through exploitation of data locality, and {\em network-level} improvements in the overhead of remaining communications through optimizing the network load.

Mapping-level communication reduction is an important optimization stage: error correction is expensive, and a reduced operation count yields multiplicative benefits. First, fewer operations must be protected against errors, and second, those that do need to be protected can afford a weaker form of correction since the overall program is smaller. We apply the mapping approach of~\cite{heckey} to the Multi-SIMD architecture, which reduces unnecessary teleportations between regions. Network-level optimizations on the Multi-SIMD architecture consist of pipelining EPR distributions to avoid high congestion, discussed in Section~\ref{sec:discuss}. The next section will discuss novel mapping- and network-level optimizations in the context of braids on the tiled architecture.

\section{Optimized Braiding}\label{sec:braid}
In this section, we describe a scalable yet efficient solution to the braiding problem, which is the main method of computation and communication in double-defect surface codes. Section~\ref{sec:braidflash} lays out our general method, while Sections~\ref{subsec:qubit_opt} and~\ref{subsec:braid_opt} propose mapping- and network-level optimizations. This corresponds to the optimization flow on the bottom path of Figure~\ref{fig:toolchain}.

\subsection{Braid Simulation using Message Passing}~\label{sec:braidflash}
\begin{figure*}[ht]
   \captionsetup[subfigure]{justification=centering}
\subfloat[Time-space volume of a 2-qubit logical operation\label{fig:cnot_braids:a}]{%
  \includegraphics[width=0.16\textwidth]{Figures/cnot3d2.pdf}
}
\hfill
\subfloat[$time = 0$Intialize Ancilla][$time = 0$\\Intialize Ancilla\label{fig:cnot_braids:b}]{%
  \includegraphics[width=0.16\textwidth]{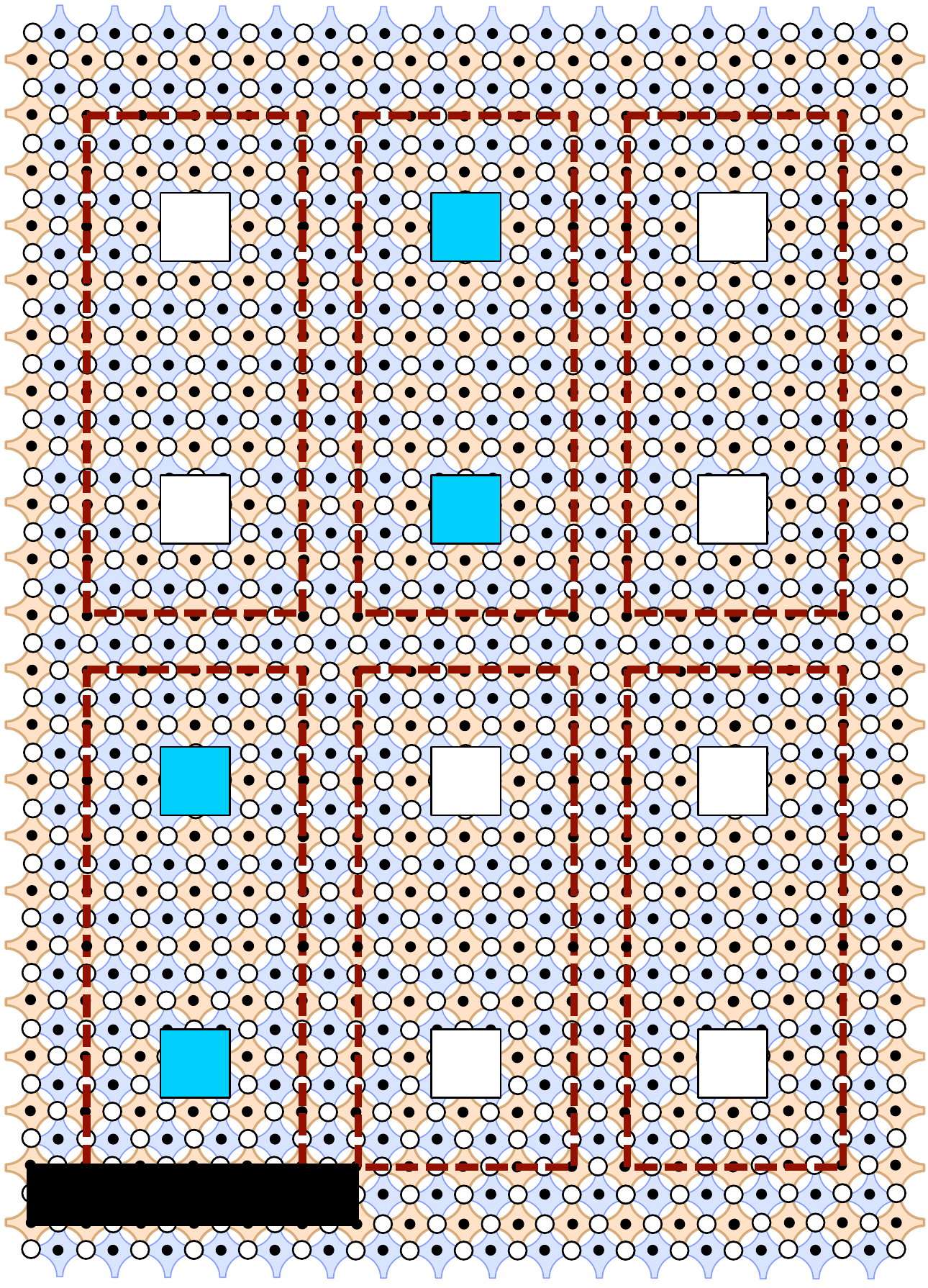}
}
\hfill
\subfloat[$time = d$Braid Part 1][$time = d$\\Braid Part 1\label{fig:cnot_braids:c}]{%
  \includegraphics[width=0.16\textwidth]{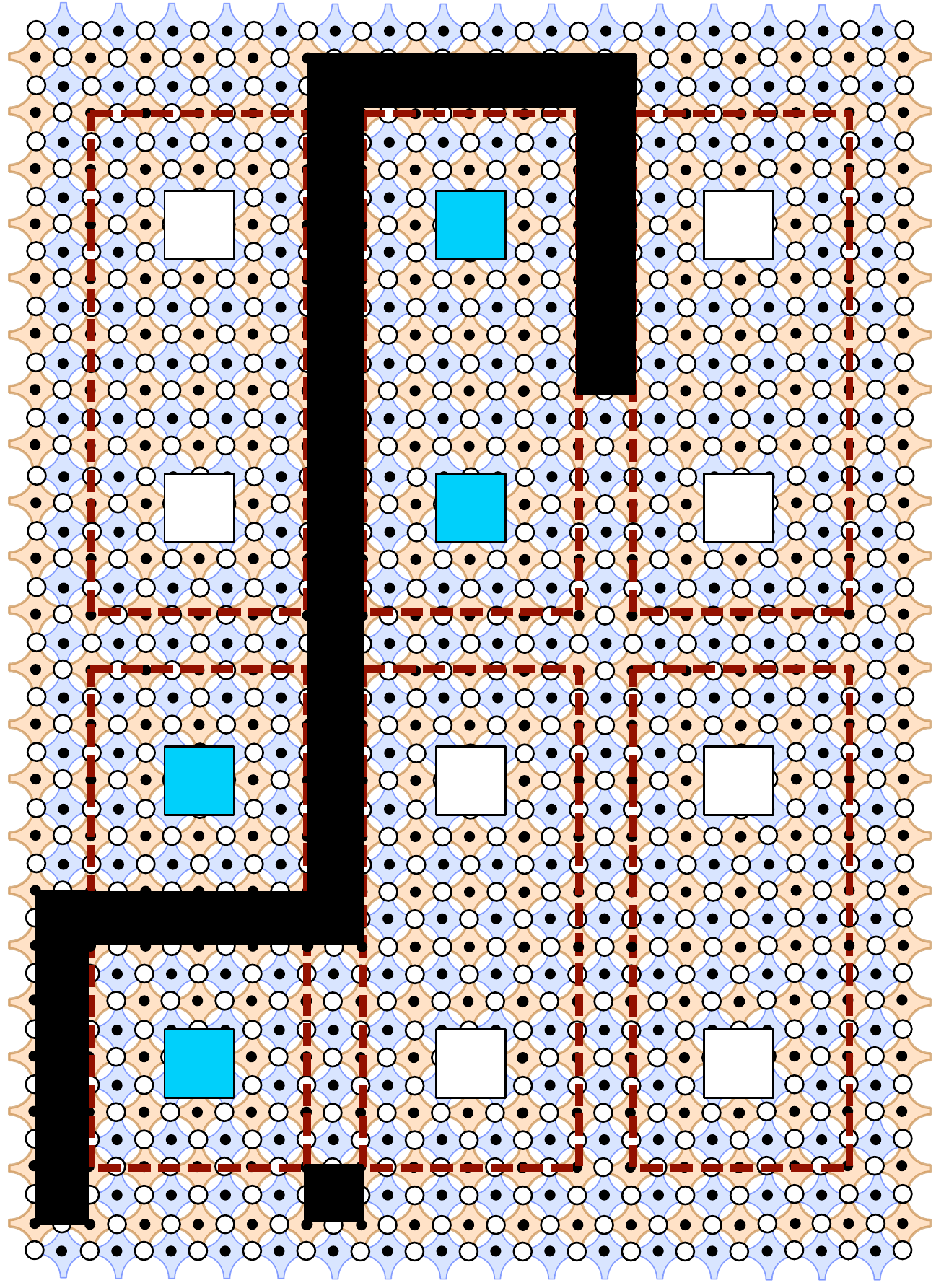}
}
\hfill
\subfloat[$time = d+1$Stabalize][$time = d+1$\\Stabilize\label{fig:cnot_braids:d}]{%
  \includegraphics[width=0.16\textwidth]{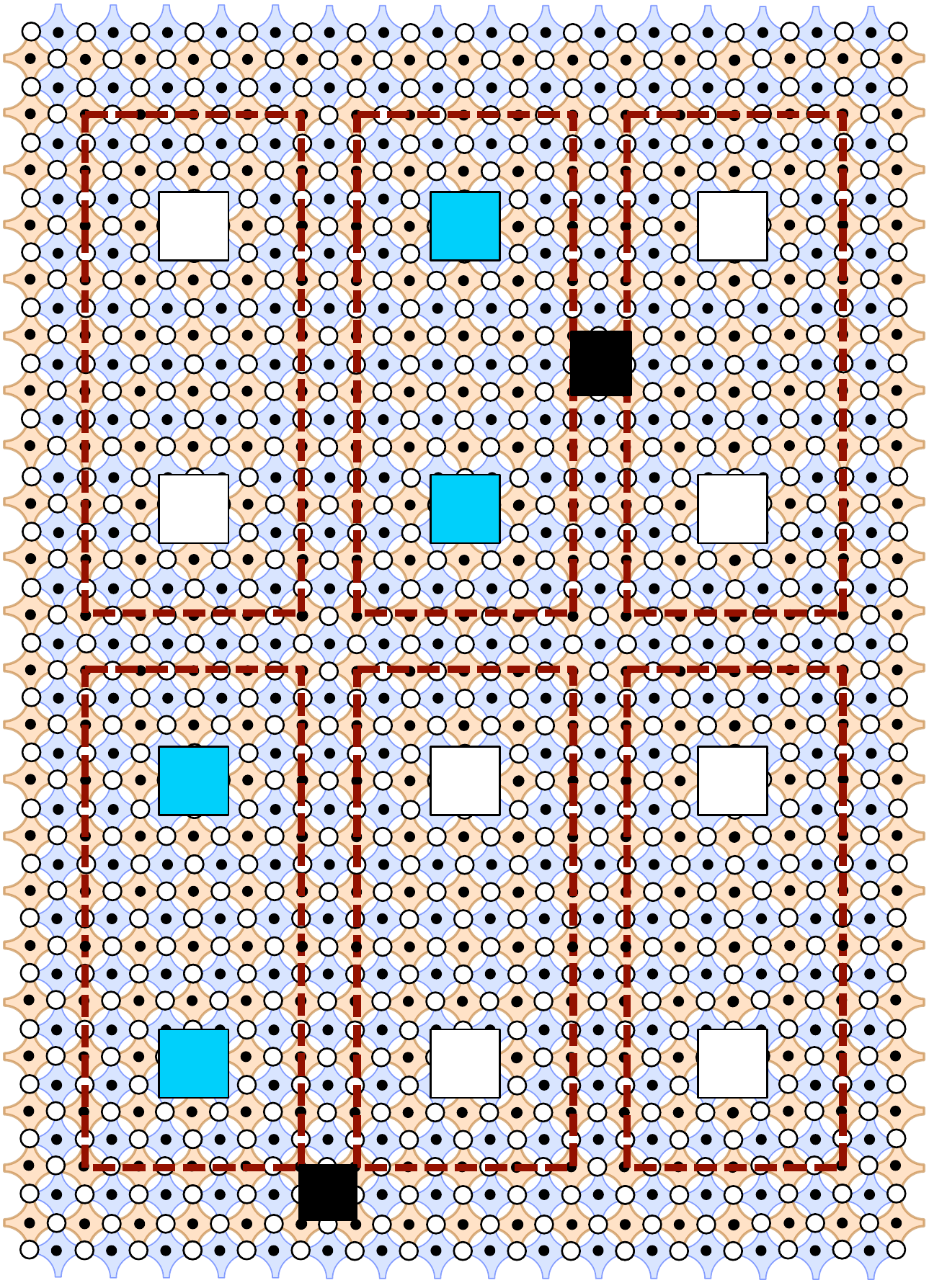}
}
\hfill
\subfloat[$time = 2d$Braid Part 2][$time = 2d$\\Braid Part 2\label{fig:cnot_braids:e}]{%
  \includegraphics[width=0.16\textwidth]{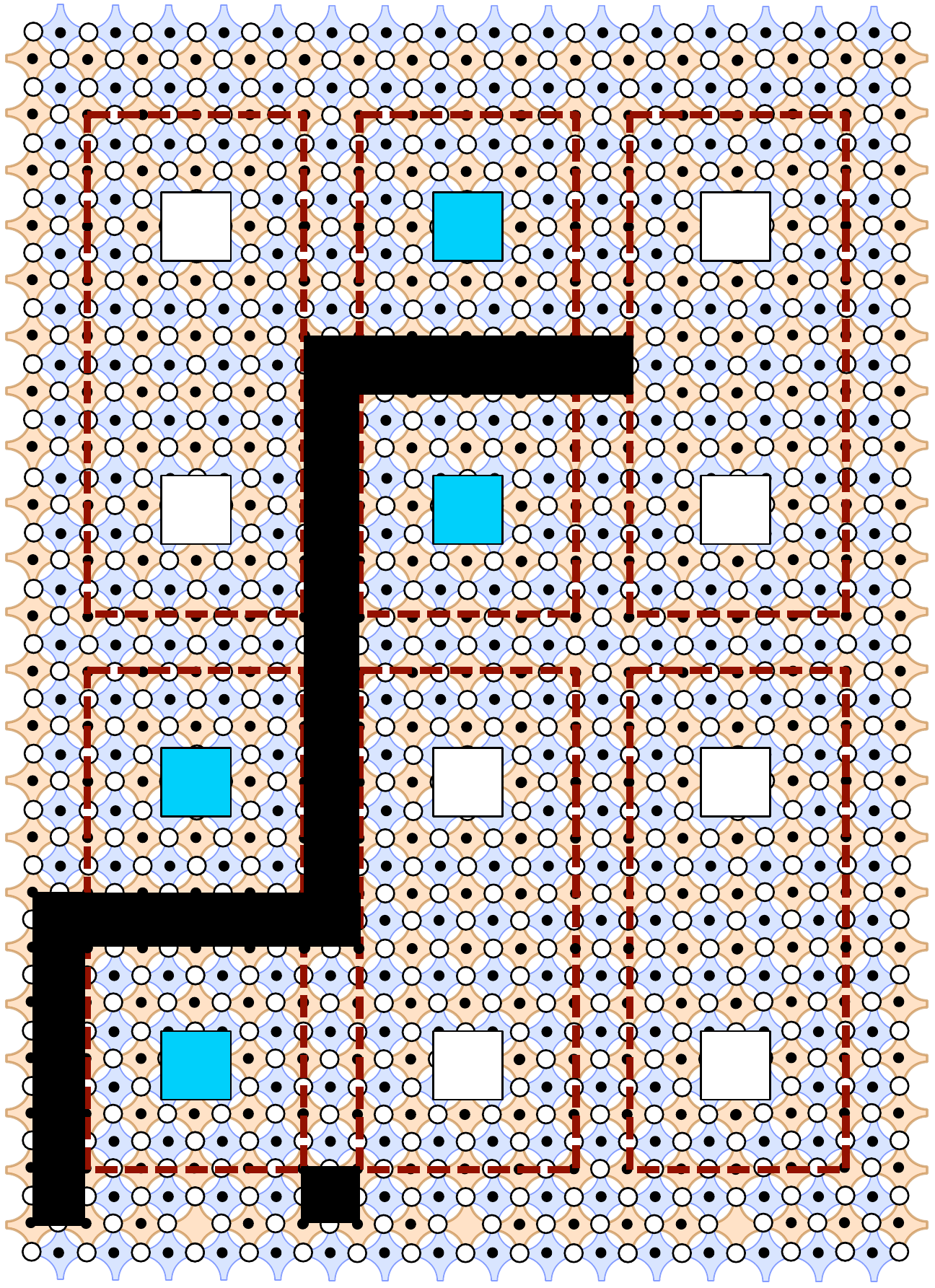}
}
\hfill
\subfloat[$time = 2d+1$Stabilize \& Measure][$time = 2d+1$\\Stabilize/Measure Ancilla\label{fig:cnot_braids:f}]{%
  \includegraphics[width=0.16\textwidth]{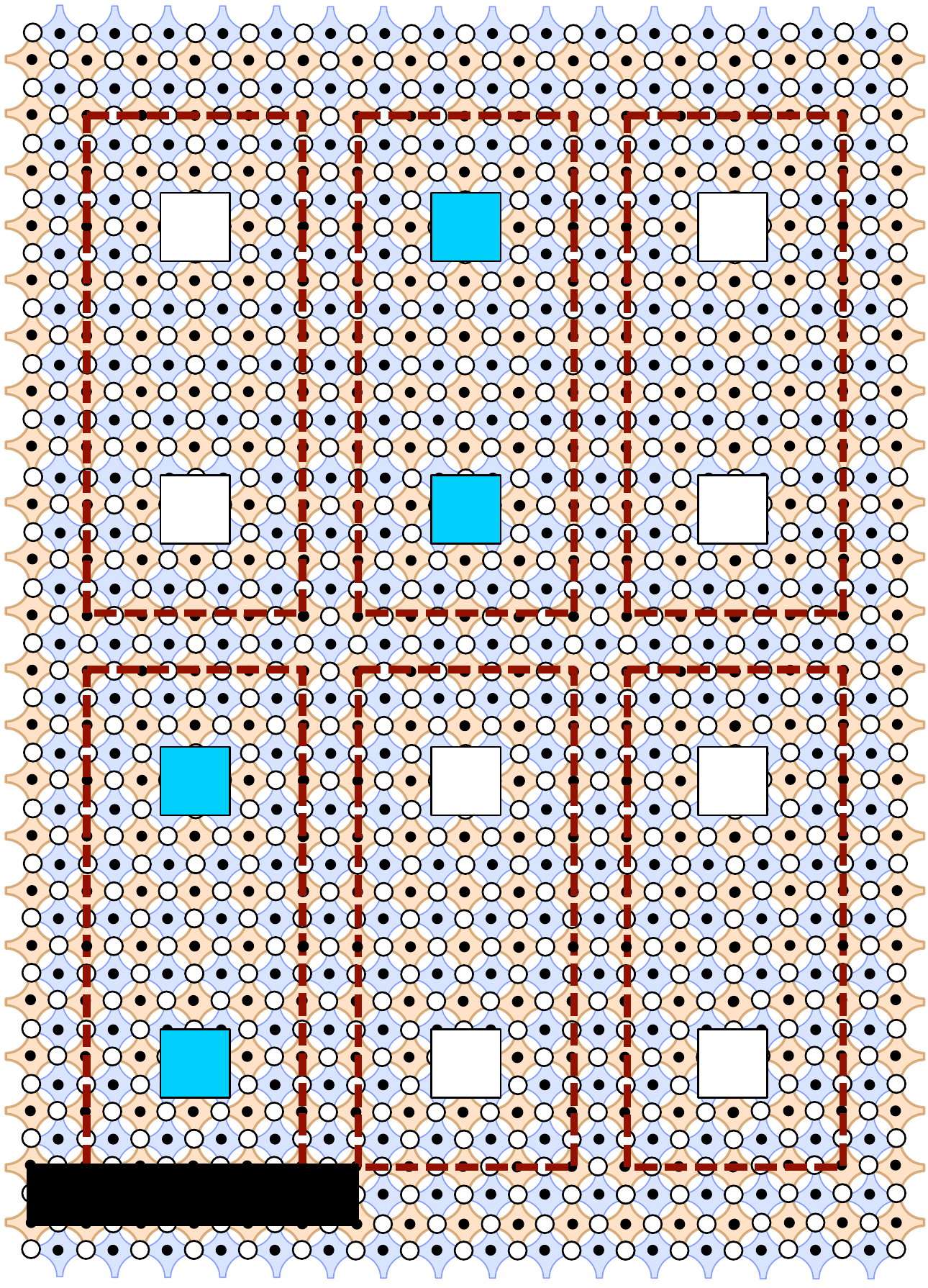}
}
\caption{Logical operation between two qubits (light blue shade for clarity) of a six-qubit setup, arranged in a 2x3 tiled architecture. (a) depicts the entire space-time volume (time goes up vertically)---the topological method used in traditional optimizations. (b-f) each represent a slice in time, called an ``event.'' From a network perspective, black defects are messages routed in the mesh, and the tile corners are routers.}
\vspace{-10pt}
\label{fig:cnot_braids}
\end{figure*}

Figure~\ref{fig:cnot_braids} shows how a simple 2-qubit logical operation may be performed between two distant double-defect logical qubits. As discussed previously, this becomes a braiding operation that amounts to topological pipes in the 3D space-time volume (Figure~\ref{fig:cnot_braids:a}). Traditionally, such volumes have been compressed by hand to decrease space and time~\cite{Fowler12,fowler2012bridge,devitt2013mequanics}. For example,~\cite{mequanics} creates a game in which players try to manually compress 3D ``pipes'' using topological transformations. Although these techniques achieve very good results on small problems, they are difficult to scale.
Instead, we propose an automated braid synthesis and optimization approach that is much more scalable, and yet achieves good performance. We do this by tracing the state of the 2D lattice over various vertical time slices in the 3D volume. As such, the problem is reduced to simulating a mesh network, with braids as messages in this network.
Figures~\ref{fig:cnot_braids:b}-\ref{fig:cnot_braids:f} show how the 3D volume can be broken down into five snapshots of braid opening and closing (i.e. message passing). We call each stage an ``event''.

This translation of the problem is an overconstraining, since not all possible topological transformations can be captured in this way (e.g. contrary to network messages, 3D topological pipes can be morphed back in time too). However, we demonstrate that with appropriate heuristics, we can reclaim much of the efficiency of optimal solutions, with the advantage of staying scalable (Sections~\ref{subsec:qubit_opt} and~\ref{subsec:braid_opt}). Our approach relies on performing dynamic routing for a static problem. The reason network routing works well here is that our applications have parallelism that is not too high and not too low. That is, we can afford to overconstrain the problem and still resolve most conflicts since there aren't too many, but we can't be too naive about it. Manual approaches try to solve every conflict as optimally as possible, but that is not necessary in practice and not scalable. In fact, our tools are what allow us to obtain these insights. Since we replay the dynamic schedule as a static one at execution time on the quantum computer, we need not worry about deadlock and livelock.  We only have to use heuristics to find a deadlock- and livelock-free solution at compile time.  This is much easier than guaranteeing deadlock- and livelock-freedom at runtime in truly dynamic schemes.

Braids differ from conventional messages in several ways: (a) braids travel $n$ hops all the way from source to destination in one cycle, (b) some braids have to remain stable for $d$ (code distance) cycles to stabilize syndrome measurements, (c) routers cannot buffer braids and (d) virtual channels cannot be used as braids cannot physically use the same channel. For these reasons, we use a circuit-switched network. The $n$ hop/cycle property means that a braid claims the route's resources (nodes and links) at once when opened, and releases them when closed. We use a greedy approach of trying to place as many braids as possible, in order to reduce overall cycles.

In our braiding algorithm, we maintain a ready queue of operations whose dependencies have been met, and execute as many of them as possible in each cycle.  Each event pertains to an open or close braid, and has a timer corresponding to how long the braid should remain in case error syndromes need to be extracted. To improve forward progress in a busy network, we add route adaptivity to a dimension-ordered route and a drop/re-inject mechanism, both after certain timeouts.  Note that we are using a dynamic mechanism (network routing) to find a {\it static schedule} (which will be replayed during the execution of our quantum program).  This means that our protocols do not need to be deadlock- or livelock-free.  If we encounter lack of forward progress, our mechanisms just backtrack and try again. There will be no cost at execution time, since failed schedules are not recorded and used.

\subsection{Optimizing Qubit Arrangement}\label{subsec:qubit_opt}
The first step to reduce braid contention is applying mapping-level optimizations to the placement of qubit tiles on the 2D mesh. Our goal is to map logical tiles which interact frequently close to each other. Specifically, the optimized arrangement of qubit tiles attempts to minimize the sum of Manhattan distances between pairs of tiles involved in non-local, braiding operations. We do this through iterative calls to a graph partitioning library, METIS \cite{metis}, to separate the qubits (each represented as a vertex on a graph of qubit interactions) into two partitions, such that the weight of crossing edges is small. Relative to a naive arrangement of qubits, the optimized qubit arrangement reduces the lengths of braids, hence reducing the chance of braid collisions.

\subsection{Optimized Braid Priorities}\label{subsec:braid_opt}
\begin{itemize}
  \setlength\itemsep{0em}
\item {\em Policy 0}: No optimization. All operations and events in program order.
\item {\em Policy 1}: Interleave: Allow individual events to be interleaved, but keep operations in program order.
\item {\em Policy 2}: Interleave + layout: Optimize initial qubit layout for interaction distances.
\item {\em Policy 3}: Interleave + layout + criticality: Sort operations by highest criticality first.
\item {\em Policy 4}: Interleave + layout + length: Sort braids by longest first.
\item {\em Policy 5}: Interleave + layout + type: Sort by closing braids first, opening braids next.
\item {\em Policy 6}: Combines Policy 1--5: Interleave, optimize initial layout, sort closing braids first, sort by criticality, sort short-to-long for highest-criticality braids, sort long-to-short for lower criticality braids. 
\end{itemize}

\begin{figure}[h]
  \centering  
  \includegraphics[width=\columnwidth]{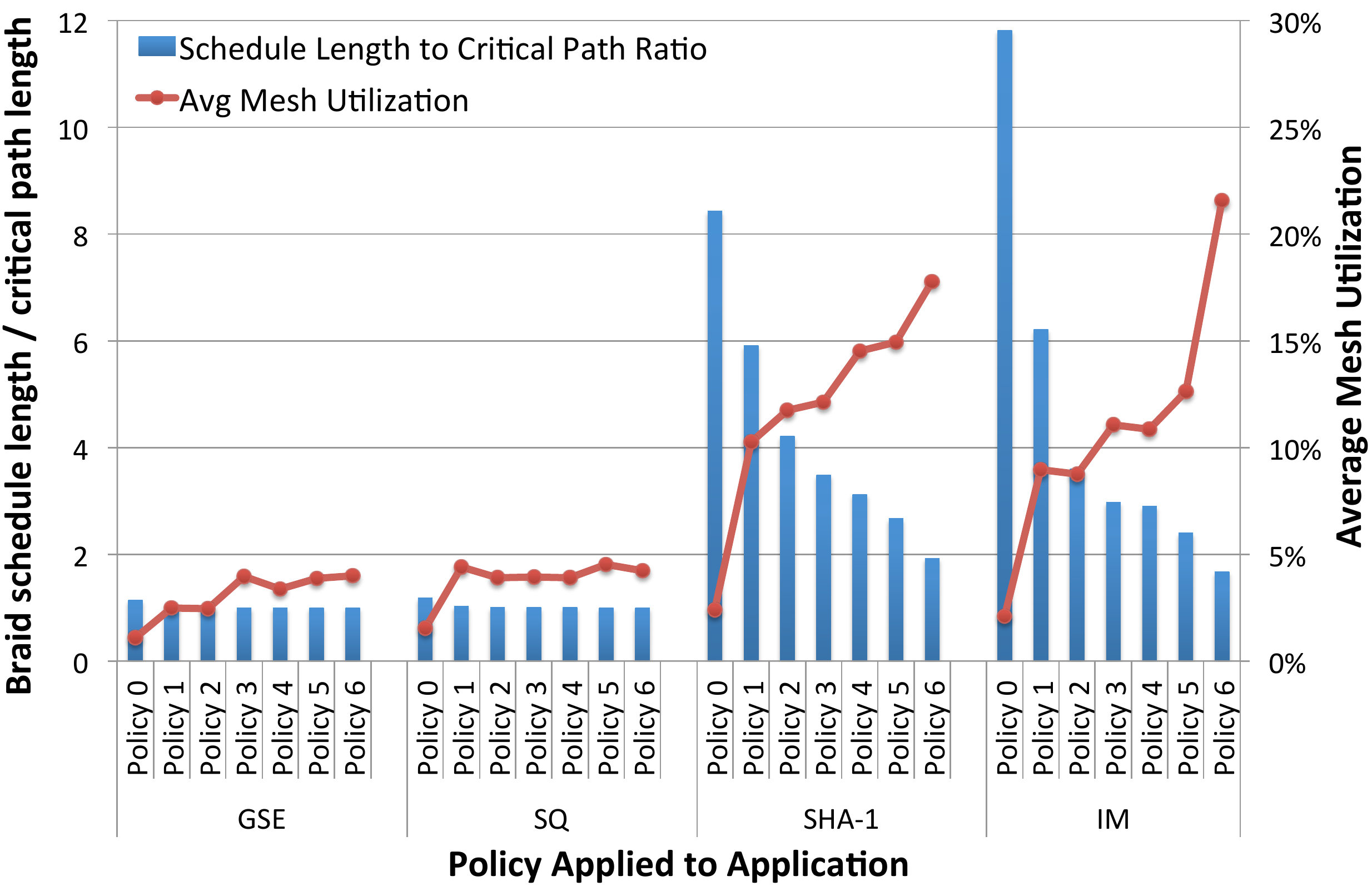}
  \caption{Braid simulation results for the double-defect surface code. Blue bars show schedule length can be reduced by up to ${\sim}7X$, to within ${\sim}70\%$ of the critical path for highly-parallel applications (SHA-1 and IM), where risk of contention is high. Serial applications (GSE and SQ) already achieve close-to-critical-path schedules. Red curves show the increase in network utilization when using these policies, up to about 22\%. These improvements are achieved through interaction-aware qubit placements and priority-aware braid placements.}  
  \label{fig:braidflash}
\end{figure}

In our braiding algorithm, when multiple braids are eligible to be scheduled but not all of them fit on the mesh, the choice of priorities can have a significant impact on performance. We have devised heuristic policies to prioritize ``more important'' braids. It must be noted that such policies are only essential in highly-parallel applications (SHA-1 and IM), where there is an initially large discrepancy between the schedule length and critical-path length, due to communication conflicts. Low parallelism reduces the need for interference optimization from the start.

The metrics we use to assess priority are the criticality of the braid (how many future operations depend on it), its length, and its type (whether it is a closing or opening braid). Our prioritization policies are summarized below:

Policy 1 (event interleaving) allows for multiple braids to progress concurrently and at different rates. Policy 2 is the same layout localization heuristic of Section~\ref{subsec:qubit_opt}. To these, Policies 3 through 5 each add one important metric by which the importance of a braid may be judged: criticality, length, type. Policy 6 combines all of the metrics above. Here, we give higher priority to those braids that are closing rather than opening (so they can release network resources), and those that have higher criticality (to remove the bottleneck). If two braids have the same criticality, we decide based on length: shorter braids for the most critical operations, because we want to accomplish as many as possible, and longer braids for lower-criticality operations, because we want to conclude the toughest braids ahead of time.

Figure~\ref{fig:braidflash} shows our results. The blue bars (associated with the left vertical axis) show how the above prioritization policies improve performance by reducing the gap between the braid schedule lengths and critical path lengths. 
Our results show that evaluated individually, braid type causes the largest improvement, while improvements due to length and criticality are smaller.

The most successful policy, Policy 6, is able to reduce schedule length by up to ${\sim}7X$, from 12X longer than the critical path, to within 70\% of it. The red curves (right vertical axis) shows network utilization rate (i.e. percentage of busy links). It demonstrates that better prioritization results in an almost 8-fold increase in network utilization---up to about 22\%. This is an acceptable range for this highly scalable approach, comparable to similar circuit-switched networks.

\section{Results}\label{sec:results}
In the context of our microarchitectures, technology, and tools, we can now compare our QEC methods for each benchmark application.

We begin by quantifying some absolute values for the number of qubits and time required to run a quantum application, and then proceed to evaluate the effect of QEC choice on performance. These results can help determine the most suitable type of error correction and microarchitecture for a particular application ($p_L$) on a particular physical technology ($p_P$).

\subsection{Scaling Effects on Space and Time}
\begin{figure*}[t]
\subfloat[Time\label{fig:abs:a}]{
  \includegraphics[width=0.47\textwidth]{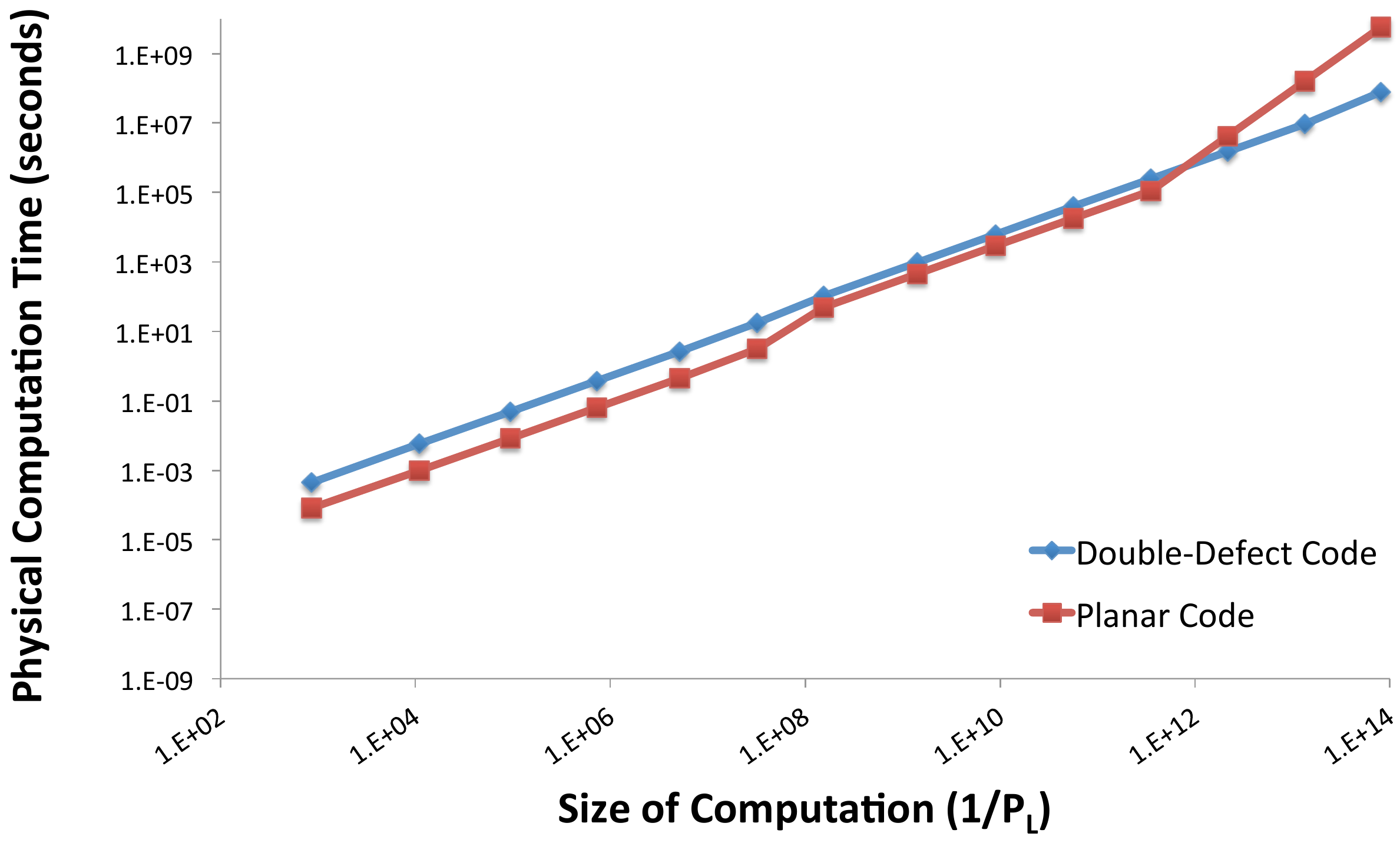}
}
\hfill
\subfloat[Qubits (Space)\label{fig:abs:b}]{
  \includegraphics[width=0.49\textwidth]{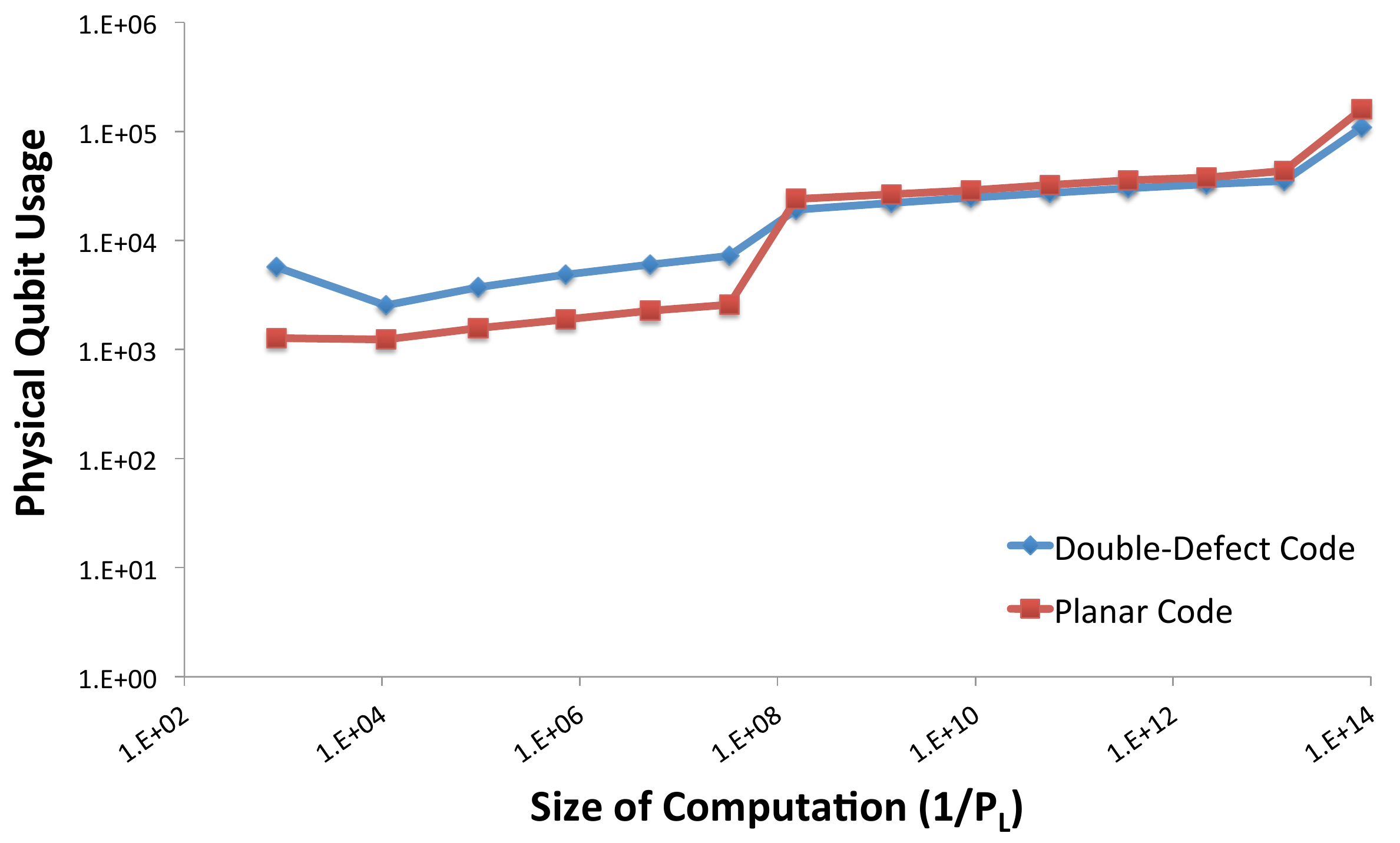}
}
\caption{Absolute resource usage in (a) time and (b) space to run error corrected SQ applications of varying sizes. In these simulations we assume $p_P=10^{-8}$ and single-qubit operations are 10X faster than 2-qubit operations.} 
\label{fig:abs}
\end{figure*}
Figure~\ref{fig:abs} shows concrete values for the number of qubits and amount of time needed to execute a fully-error-corrected SQ application. The graphs are plotted for various input problem sizes, each of which directly determines the size of computation (inversely proportional to target logical error rate ($p_L$)). Superconductors are fast, and we see that small instances of the problem can execute in under one second. Increasing the problem size, however, can greatly increase time of computation. 
The number of qubits does not rise as sharply, but it still illustrates the scaling needs of future computers: in order to run even modest problem sizes, around 1000 qubits are needed. The main increases in qubit usage occur when the code distance ($d$) must be increased to support larger computations. Other applications exhibit similar scaling trends.

Although Figure~\ref{fig:abs} is useful for obtaining a sense of absolute overheads, it does not depict the difference between different encodings well---the lines are close to each other due to the logarithmic axes. Next, we look at the ratios of double-defect to planar resource usage.

\subsection{Effects of Encoding}
\begin{figure}[ht]
\subfloat[SQ Application (Serial)\label{fig:app_dep:a}]{
  \includegraphics[width=0.48\textwidth]{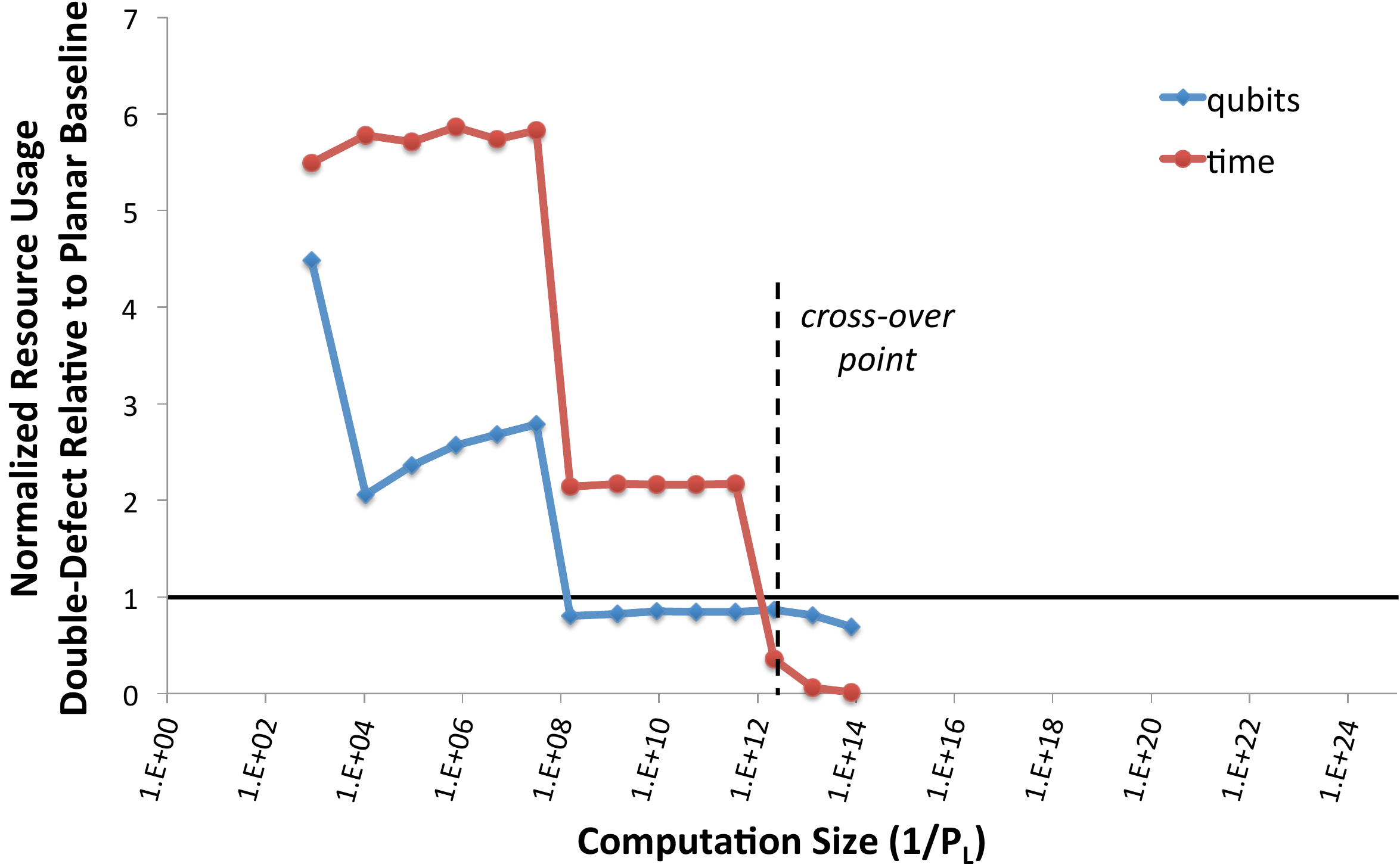}
}
\hfill
\subfloat[IM Application (Parallel)\label{fig:app_dep:b}]{
  \includegraphics[width=0.48\textwidth]{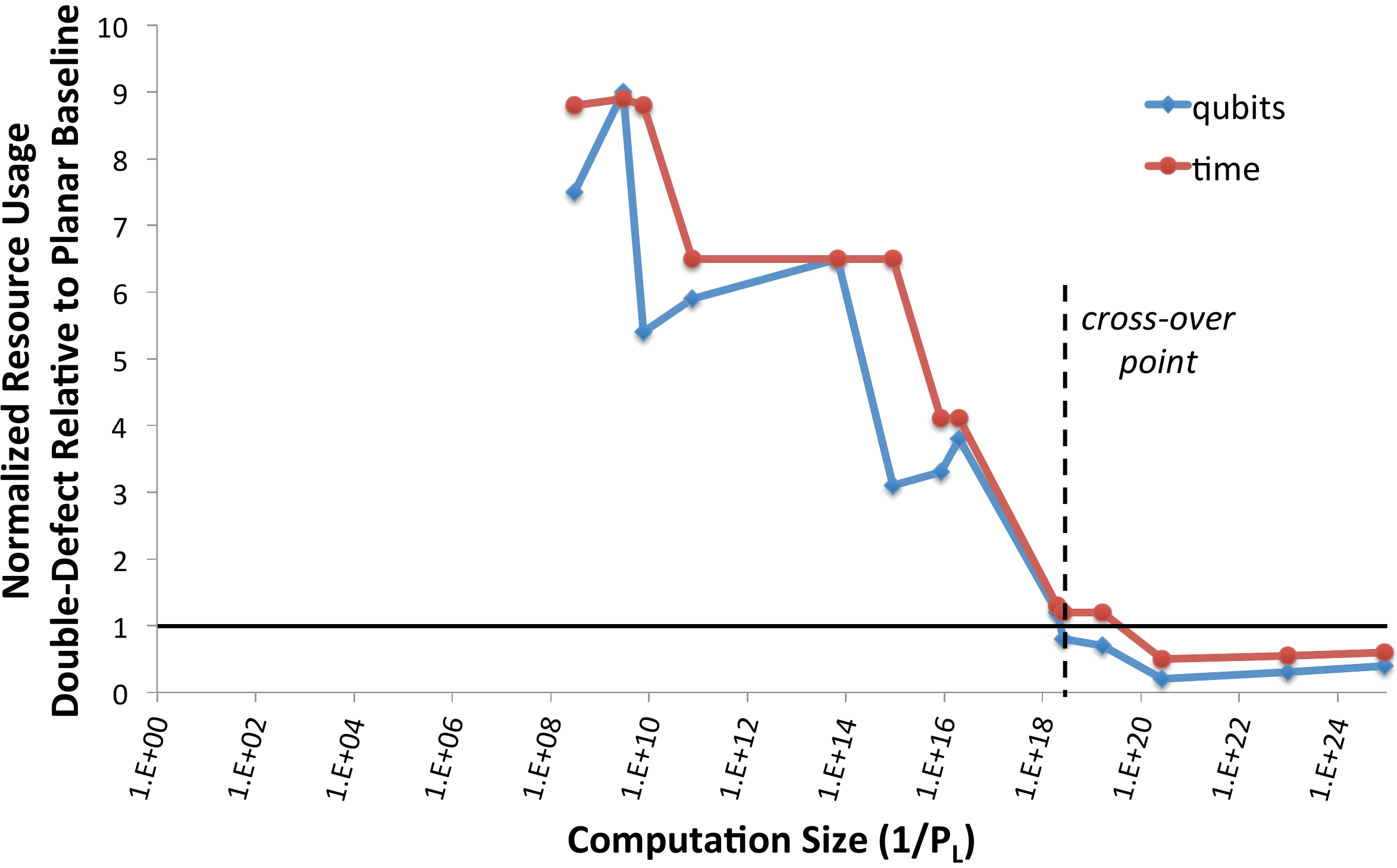}
}
\caption{Comparing resource usage in double-defect and planar codes, for a range of problem sizes in the (a) SQ and (b) IM applications. Lower is better, and overall preference is given to the QEC method that has a smaller $qubit \times time$ product. In both (a) and (b), planar codes are better at smaller sizes but at some cross-over point, double-defect codes become better. However, the cross-over point occurs at a much larger computation size for IM, compared to SQ. This is due to high braid congestion in the highly parallel IM app, which causes planar (teleport-based) codes to remain better for longer. Figure shown for a technology with error rate $p_P=10^{-8}.$} 
\vspace{-10pt}
\label{fig:app_dep}
\end{figure}

We now explore the differing overheads when using planar and double-defect encodings in two applications, SQ and IM, specifically chosen for their parallelism characteristics (Figure~\ref{fig:app_dep:b}). Our metric is resource usage (qubits and time), normalized to planar codes as a baseline.

SQ is largely serial, whereas IM is highly parallel. In both applications, when the computation size is small, planar codes fare better. This is due to the smaller size of planar lattices. However, once the computation size exceeds a certain amount (indicated as the ``cross-over point''), double-defect codes become better, due to the efficiency of braids compared to slower swaps. Favorability cross-over occurs where the space-time ratio ($qubits\times time$) crosses 1.

Comparing Figures~\ref{fig:app_dep:a} and~\ref{fig:app_dep:b} we find that the cross-over point in IM occurs much later than SQ (at a larger computation size). This is directly due to the high parallelism of the IM application. Parallelism causes braid congestions in double-defect codes, but the EPRs in planar codes can still be pipelined in a way to avoid congestion. Therefore, planar qubits remain better for longer. Furthermore, the Multi-SIMD architecture of planar codes supports data and instruction parallelism, again improving the performance of planar codes which can use parallel bitwise operations in this architecture. Note that these results pertain to the fundamental difference in braiding versus teleportation, and even though we have evaluated them in the context of specific architectures, our observations apply to other proposed architectures~\cite{musiqc,qla} (which, in fact, are more favorable to planar codes).

\subsection{Sensitivity Analysis}~\label{sec:sensitivity}
\begin{figure}[!ht]
  \includegraphics[width=0.5\textwidth]{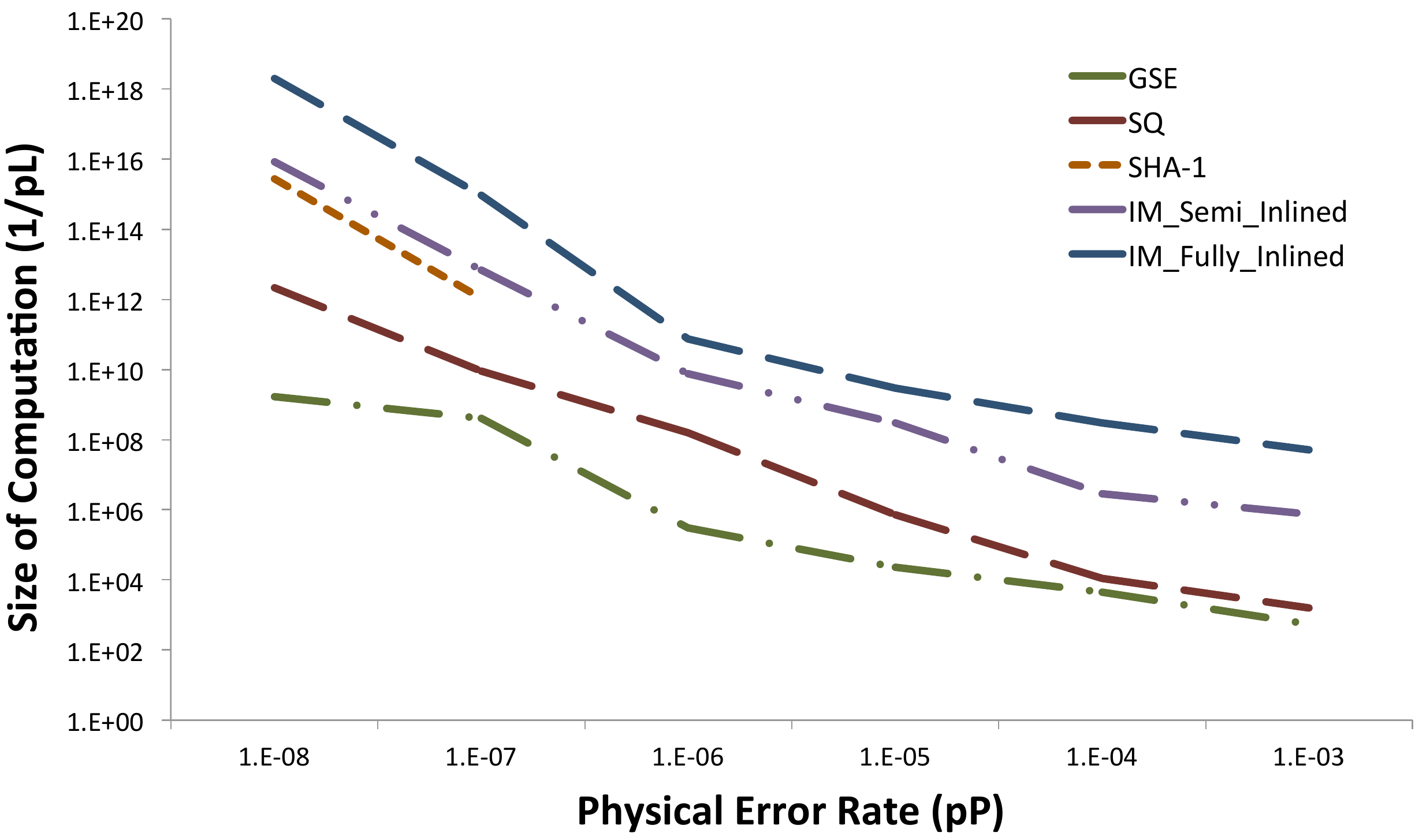}
\caption{Comparison of double-defect vs. planar surface codes, across the range of possible physical and logical error rates. Higher y-axis value (up) means more logical operations in the application. Higher x-axis values (right) mean faultier technology. Design points under the curves work better with planar codes. Each line delineates the cross-over boundary for a particular application. The curve is higher for more parallel applications (SHA-1, IM), since congestion hurts braids more.} 
\vspace{-10pt}
\label{fig:comp}
\end{figure}

Having established how a cross-over occurs from the favorability of planar codes to the favorability of double-defect codes, we can now plot this point for a full sweep of physical and logical error rates for all applications. Figure~\ref{fig:comp} shows this graph. Each line on this graph is a collection of cross-over points, showing their sensitivity to changing physical error rates from $p_P=10^{-8}$ (future optimistic) to $p_P=10^{-3}$ (current~\cite{sheldon2016characterizing,sheldon2016procedure}). For each application, the line demarcates the boundary of designs where planar codes should be used (below) and where double-defect codes should be used (above).

Such lines are application- and technology-dependent, due to the entirely different models of computation (bitwise vs. braids) and communication (teleportation vs. braids) in these codes, which must be simulated in order to yield accurate contention rates. We observe that boundaries are generally higher for more parallel applications, suggesting that these applications would benefit more from using planar codes. We have used two variations of the IM application---with medium and maximal inlining. More code inlining creates more parallelism, consistent with the upward boundary movement.

In summary, our results suggest that although double-defect codes have been considered the standard method of error correction on superconducting computers before~\cite{FowlerSurface,hifi_supercond_1}, subtleties related to communication contention and application parallelism may alter this view. As device error rates continue to improve (left in Figure~\ref{fig:comp}), a shift to planar encoding may be warranted (bigger area under the curves).

\section{Discussion}\label{sec:discuss}

\subsection{Pipelined EPR Distribution}\label{subsec:smoothing}
In the communication technique referred to as teleportation, the physical interconnection network does not directly move the data qubits themselves, but instead carries EPR qubits that facilitate the communication of entangled qubits at a distance. 

Because of the delay-tolerant nature of the distribution of EPRs, they are not bound by stringent data dependencies, and they can be prefetched at arbitrary points in time. Our goal is to achieve ``just-in-time'' distribution by smoothing the network load to reduce congestion. This can have favorable effects on both space and time: the number of EPR qubits will be reduced as they are consumed (and recycled) shortly after being introduced to the network; latency will be reduced since EPRs are present when they are needed, and hence do not stall teleportations.

Given the specialized nature of quantum applications, we have perfect static knowledge of when each EPR will be needed. Hence, walking the dependency graph, we use look-ahead windows to anticipate usage points, and launch their communication with appropriate lead time. Such approaches achieve good performance across all applications, with up to~${\sim}24X$ savings in qubit cost and only a maximum of~${\sim}4\%$ extra latency. 
The choice of lead time (``window size'') is important to achieve just-in-time distribution. Smaller window sizes cap qubit usage at the expense of starving data qubits in need of teleportation. In contrast, large windows release more EPRs into the network than necessary. Suitable window sizes depend on application size, but degree of application parallelism has little effect, since ancillas do not follow regular data dependencies.

\subsection{Alternative Communication Methods}\label{subsec:latticesurgery}
While this paper has discussed teleportation and braiding as two main communication methods, recent work has investigated {\em lattice surgery}~\cite{LatticeSurgery} as a hybrid scheme that combines the low qubit overheads of the planar code with nearest-neighbor-only interactions. Communication in lattice surgery occurs using {\em merge} and {\em split} operations: two adjacent planar encoded qubits are merged by turning on syndrome measurements at their connecting boundary, creating a larger continuous plane. Similarly, turning those syndromes off again will split the two planes into their original form. Repeated applications of these operations in a chain can cause distant planes to interact. Optimal lattice surgery is an NP-hard problem~\cite{latticeNP}, and scalable heuristics remain as an area of future work. Crucially in the context of this study, the chain of merges and splits does not have the benefits of braids (fast movement) nor teleportation (prefetchability). Therefore, we instead focused on two more promising forms of communication.

\section{Conclusions}~\label{sec:conclusion}
This paper has focused on optimizing qubit communications as a major source of latency in quantum computers. Focusing on surface code error correction in superconducting qubits, the leading candidate to implement a future quantum computer, we investigate teleportation-based and braid-based communications. We present optimization algorithms that achieve near-critical-path performance, while staying scalable to problem sizes of more than $10^{20}$ operations. 
Our design-space exploration allows us to plot favorability curves such as Figure~\ref{fig:comp}, which is useful as a guide in QC design. For example, it shows us that for near-term superconductor error rates of $10^{-4}$--$10^{-3}$, planar encoding is better for serial applications of shorter than ${\sim}10^4$ logical operations and parallel apps of shorter than ${\sim}10^{8}$ logical operations.
We believe the large gains from running a quantum application is such that each real quantum machine will likely be dedicated to solving a specific task; e.g. breaking large encryption keys or finding the ground state of large molecules. It is therefore suitable to pick design parameters based on the intended application. Similarly, we have shown that the physical device characteristics (i.e. error rate) must play a role in choosing suitable error correction codes and microarchitectures. This paper therefore argues for a co-design of applications, microarchitectures and physical hardware in building future quantum computers.


\bibliographystyle{ACM-Reference-Format}
\bibliography{references}

\end{document}